\documentclass[
  twocolumn,
  prb,
  preprintnumbers,
  floatfix,
  superscriptaddress,
  citeautoscript,
]{revtex4}

\usepackage{physics}
\usepackage{amsmath}
\usepackage{amssymb}
\usepackage[dvipsnames]{xcolor}
\usepackage{graphicx}
\usepackage{tabularx}
\usepackage{subfigure}
\usepackage{dcolumn}
\usepackage{times}
\usepackage{comment} 
\usepackage{ulem}

\newcommand{\comnt}[1]{}

\newcommand{\sect}[1]{Sect.~\ref{#1}}

\newcommand{\tab}[1]{Table~\ref{#1}}

\renewcommand{\vec}[1]{\ensuremath\boldsymbol{#1}}
\renewcommand{\epsilon}[0]{\varepsilon}

\def\biblio{
\bibliographystyle{unsrt}
\bibliography{ourref}
}

\usepackage{subfiles}

\usepackage{xr}

\begin{document}

\def\biblio{}

\title{  Ab initio calculations for void swelling bias in  $\alpha$ and $\delta$-plutonium  }

\author{Babak Sadigh}
\email{sadigh1@llnl.gov}
\author{Per S\"oderlind}
\author{Nir Goldman}
\author{Michael P. Surh}
\email{surh1@llnl.gov}

\affiliation{
  Lawrence Livermore National Laboratory,
  Physical and Life Sciences Directorate,
  Livermore, CA, 94550
}

\date{\today}
\begin{abstract}

Void swelling can develop in materials under persistent irradiation when non-equilibrium vacancy and self-interstitial populations migrate under sufficiently asymmetric interaction biases.
In conventional metals, the propensity is determined to first approximation by comparing point-defect relaxation strains.
We thus present DFT-based calculations of structures and formation energies and volumes of point defects in the $\alpha$ and the $\delta$-phases of plutonium. 
We discuss pros and cons of various levels of electronic structure theory: spin polarization, spin-orbit coupling, and orbital polarization.
Our results show that lattice defects in $\delta$-Pu, in contrast to most fcc metals, have surprisingly small formation volumes. Equally unexpected are the large defect formation volumes found in the low-symmetry $\alpha$-Pu phase. Both these unusual properties can be satisfactorily explained from defect-induced spin/orbital moment formation and destruction in the Pu phases. Surprisingly, the point defects in $\alpha$-Pu are found to induce far larger transformation of the local electronic structure than in $\delta$-Pu. When we use the calculated defect properties to estimate the classic void swelling bias in each of the phases, we find it to be unusually small in $\delta$-Pu, but likely much larger in $\alpha$-Pu. Hence, swelling rates and mechanisms can diverge dramatically between the different phases of Pu. Especially in the transient regime before formation of large defect clusters, the swelling rate of $\alpha$-Pu can reliably be expected to be much larger than $\delta$-Pu. However, accurate forecasts over longer times will require the conventional void-swelling theory to be modified to handle the complexities presented by the different Pu phases. As a case in point, we show possible anomalous temperature dependence of vacancy properties in $\delta$-Pu, caused by entropic contributions from defect-induced spin-lattice fluctuations. Such complications may affect defect-defect interactions and thus alter the void swelling bias. 
\end{abstract}

\biblio

\maketitle

\section{Introduction}
\label{sect:introduction}

Pu-based materials are inherently subject to radiation damage that can adversely affect their properties over time.
Spontaneous $\alpha$-decay deposits both U and He impurities and displaces host atoms from their lattice sites in collision cascades.
It leaves behind a non-equilibrium population of point defects and clusters that feed compositional, microstructural, and macroscopic changes over the long term. Given sufficient time, the defect populations and microstructure co-evolve via biased defect diffusion as well as defect collisions and reactions at defect sinks in the lattice. For Pu metal in particular, these long term radiation aging effects  may encompass: (i) He bubble formation\cite{density:2000,ARSENLIS:2005,bubble:SCHWARTZ:2005,CHUNG:2006,CHUNG:2007,JOMARD:2007,BERLU:2008a,DREMOV:2008,DREMOV:2011,JEFFRIES:2011,LI:2011,AO:2012,KARAVAEV:2014,ROBINSON:2014,CHUNG:2016,WU:2017,KARAVAEV:2017},
(ii) metastability and possible phase transformation \cite{GaStable:SADIGH:2005,review:SCHWARTZ:2007,VALONE:2007,review:SCHWARTZ:2009,DREMOV:2013,LI:2017}, and (iii) void swelling \cite{GSWas:2016,review:WOLFER:2000,review:BOOTH:OLIVE:2016}. The possibility of void swelling caused by self-irradiation is the focus of this paper. 

Swelling behavior has been observed and studied in depth in reactor materials.
It arises from incomplete mutual annihilation of the relict, post-collision-cascade lattice defects and their subsequent net segregation to pre-existing sinks in the microstructure. 
In the conventional theory of void swelling in metals, the long-range interaction that drives segregation originates from coupling of the (mobile) point defect relaxation strains with the local elastic stress field.
That stress field is primarily induced by dislocations \cite{review:WOLFER:2000}, and as a consequence the diffusing defects preferentially drift towards the dislocations. The relaxation strain is also commonly much larger for self-interstitials (SI) than for vacancies (V) and so in turn the SI diffusion bias is stronger. Hence, even if V and SI are generated at the same rate by the collision cascades, the SI are absorbed by the dislocations at higher rate, resulting in climb. This in turn leads to an imbalance in the populations of the free V and SI. Void swelling comes about by accumulation of the excess residual vacancies into cavities, typically via inhomogeneous nucleation at helium bubbles.\cite{HARKNESS:LI,BRAILSFORD:BULLOUGH,WIEDERSICH:1972,RUSSELL:1978,HEALD:SPEIGHT:1974,WOLFER:ASHKIN:1975a,WOLFER:ASHKIN:1975b,GSWas:2007}

In conventional metals, the phase diagrams are dominated by one or two simple crystal structures such as fcc, bcc or hcp; their metallic bonding is well-described by Hohenberg-Kohn-Sham density functional approximations.
The relevant defect formation energies and relaxation volumes can thus readily be obtained from {\it ab initio} total energy calculations within the density-functional theory (DFT). Plutonium metal, however, is far from conventional and it was long not obvious how to best model its electronic structure in order to predict its complex phase diagram with reasonable accuracy from first principles. The 5$f$ electrons in Pu are on the knife's edge between localized and delocalized character; indeed, the material is often suspected of lying near a quantum critical point. Crystalline plutonium undergoes six structural transformations as a function of temperature, which involve large volume changes. Only two of these solid phases are of interest here: the mechanically brittle low-symmetry monoclinic ground-state $\alpha$ phase, and the ductile high-symmetry face-centered cubic $\delta$ phase, which becomes thermodynamically stable between 592-724 K. The specific volume of the $\delta$-Pu phase is $25\%$ larger than $\alpha$-Pu. The inclusion of just a few atomic percent of Ga stabilizes the $\delta$-Pu structure down to room temperature. Due to slow kinetics, it took decades before it was proven that Ga-stabilized $\delta$-Pu phase is only metastable. The true equilibrium consists of the two-phase coexistence of $\alpha$-Pu phase and the Pu$_3$Ga line compound \cite{TIMOFEEVA:2001,review:HECKER:2000}. Questions inevitably remain whether spontaneous self-irradiation in the $\alpha$- and $\delta$-Pu phases may influence the kinetics towards equilibrium. This calls for a comparative study of radiation-induced defects and possible swelling behaviors of these two phases.

In order to justify the approach to modeling the electronic structure of plutonium that is undertaken in this paper, we briefly review the history of modeling and explain our current understanding of this extraordinary material. The first realistic total-energy calculations for plutonium \cite{SODERLIND:1997} assumed delocalized 5$f$ electrons in the electronic structure $i.e.$, the electrons are considered fully bonding itinerant (Bloch states) and analogous to the $d$ electrons in the $d$-transition metals. The bonding electrons are identified by the parabolic decrease of the atomic volume proceeding half-way through the $d$-transition metal series and the following parabolic increase due to the gradual filling of anti-bonding $d$ states \cite{FRIEDEL:1969}. The early part of the actinide series shows the same behavior with a parabolic volume contraction up until plutonium. The ground-state $\alpha$-Pu phase only deviates slightly from the parabolic trend. Yet at slightly elevated temperatures, the $\delta$ phase has almost 25\% larger volume, which appears completely inconsistent with the bonding (delocalized) 5$f$-electron picture. Early on it was thus concluded that the $\alpha$-phase can be modeled adequately within DFT, while the $\delta$-phase cannot.  In this initial approach, the latter phase was predicted to have much too small atomic volume and much too high energy.\cite{SODERLIND:1997}

As a contemporary solution to this dilemma, the concept of {\it partial localization} of the 5$f$ electrons was introduced in various ways \cite{PENICAUD:1997,ERIKSSON:1999,PETIT:2000,SODERLIND:2003}. The idea behind partial localization is that the manifold of 5$f$ states in $\delta$ plutonium is divided into localized and delocalized parts. These models could adequately reproduce the $\delta$ volume and even in one case produce electronic density of states consistent with experimental photoelectron spectroscopy \cite{ERIKSSON:1999}. The most widespread approaches to modeling strong $f$-electron correlations/localization in $\delta$-Pu have been via Hubbard Hamiltonian for the $f$-electrons with strong intra-atomic screened Coulomb repulsions on the order $U\sim 4$~eV. These are combined with DFT in self-consistent schemes either through the static DFT+U approach\cite{SAVRASOV:2000} or the more advanced Dynamical Mean Field Theory (DMFT) approach \cite{DAI:2003}. Nonetheless, an important drawback of all these approaches is their reliance on an external parameter $U$, which in the partial-localization picture must vary between the different phases of Pu, because they may involve very different fractions of localized 5$f$ electrons.

Concurrently, an alternative  based on itinerant 5$f$ electrons was proposed that could deal with both the $\alpha$ and $\delta$ phases in a unified manner \cite{SODERLIND:2001}. This approach consists of a straightforward application of spin-density functional theory, while in addition, the electrons are allowed to couple through spin-orbit interaction and atomic Hund's second rule coupling (absent in conventional DFT) leading to formation of spin and orbital moments. Within this approach, atomic volumes, energies, and even simulated photoemission spectra \cite{SODERLIND:2002} for $\delta$-plutonium have been accurately reproduced. While these results suggested a viable pathway to ab-initio modeling of the materials properties of Pu, their prediction of static magnetic moments in Pu \cite{SODERLIND:2001} was criticized \cite{LASHLEY:2005} because of their apparent disagreement with experimental data at the time. However, the existence of magnetic moments in plutonium, albeit fluctuating, has since been verified in neutron-scattering experiments \cite{JANOSHEK:2015}. The measured magnetic form factor from these experiments is rather well reproduced by the accompanying \cite{JANOSHEK:2015} DMFT calculations that assume strong electron correlations (Hubbard U $\sim$ 4 eV). On the other hand, the magnetic form factor was also correctly predicted years earlier within the present DFT+OP model \cite{SODERLIND:2007,SODERLIND:2015}. Therefore, the measured magnetism can be explained within two rather different models.

At present, the validity of the partial-localization picture and the exact nature of $f$-electron correlations in different plutonium phases remain unsettled. Note that in heavy actinide metals, where localization-delocalization transitions of 5$f$ electrons do undeniably occur, they are associated with large energy changes, induced by enormous (Mbar) pressures \cite{LINDBAUM:2001,HEATHMAN:2013}. In contrast, the $\alpha$- and the $\delta$-Pu phases must be very close in energy because their equilibria are only separated by a modest temperature at ambient pressure. Hence, there is clear distinction between the behavior of the 5$f$ electrons in plutonium metal and in the heavier actinide metals. Along these lines in recent years, DMFT calculations with small Hubbard U (less than 1 eV) \cite{AMADON:2016}, have shown significant success in accounting for quantum fluctuations in the Pu system. 

The DMFT approach attempts to take into account local onsite quantum fluctuations by constructing effective Hubbard Hamiltonians for the 5$f$-electron subspace through projections onto each atomic site and solving approximately by mapping each site onto an Anderson impurity problem. This is computationally very demanding. In comparison, the spin-density functional theory amended with orbital polarization the approach taken in this paper) is computationally far more expedient. Even at this level of approximation, achieving well-converged calculations for point defect properties, which require periodic supercells containing more than 100 Pu atoms is quite daunting, It is therefore currently the only technique that is computationally feasible and at the same time reasonably accurate for the comprehensive study that is undertaken in the present work.

Nevertheless, this approach remains a static mean-field approximation to a more complete theory that can account for quantum fluctuating moments. Existing DFT approximations cannot account for this possibility. Hence a closer look at the sources of error that can arise when applied to the different phase of Pu metal is warranted. The most obvious discrepancy is in the static ground-state magnetic configuration of the $\delta$-Pu phase. In the collinear spin/orbital-polarized limit, it is found to be a layered antiferromagnetic (AF) configuration consisting of ferromagnetic fcc-(001) layers, with adjacent layers having opposite spin orientations, so-called L10 pattern. It has tetragonal symmetry, and thus breaks the fcc-cubic symmetry of the $\delta$-Pu phase.  and favors a face-centered tetragonal (fct) crystal structure. The effect is weak: at the equilibrium lattice constant, the calculated c/a for the fct structure is 0.99 in SP+SO+OP-GGA, very close to cubic. This poses no difficulty for the description of the  high-temperature $\delta$-Pu phase, where thermal fluctuations could stabilize the high-symmetry phase. However, the discrepancy becomes conspicuous for the Ga-stabilized $\delta$-Pu, which is presumably cubic down to very low temperatures although a small tetragonal distortion has been inferred from neutron powder diffraction experiments.\cite{LAWSON:2000,LAWSON:2005} This 1\% effect may reflect the neglect of quantum spin and orbital fluctuations. Later in this paper, we conduct a careful analysis of the anisotropy of structural relaxations around the point defects in $\delta$-Pu. This helps to quantify the effect of this artifact on the point defect properties, showing it to be finite but small. Regardless of the bulk results, the point defect properties in both phases of Pu in this study are found to be quite anomalous. Our analysis of the origins of these anomalies, discussed below, suggest that incorporation of quantum fluctuations will likely not qualitatively change the conclusions reached in this paper.

To summarize, our preferred plutonium treatment, due to its accuracy and relative computational efficiency, is that of a spin-polarized DFT electronic structure perturbed by (for plutonium) rather significant spin-orbit coupling (SO) and orbital polarization (OP). The generalized gradient approximation (GGA) is made for the DFT electron exchange and correlation; it is known to be the best choice to date for actinides (see discussions in Ref.~\onlinecite{review:SODERLIND:2019}).
Addition of SO and OP corrections consistently enhance the predictive capability of SP-GGA for the energetics of all phases of Pu metal.  As a matter fact, it has been demonstrated that the SP+SO+OP-GGA functional can reproduce the energy ordering and structural properties of all the experimentally observed Pu polymorphs in the phase diagram quantitatively \cite{SODERLIND:2004}. Similarly good agreement has been found for phonon properties \cite{SODERLIND:2015,SODERLIND:2019b}.

It should be noted that even without spin-orbit and orbital polarization, relativistic spin-polarized DFT (SP-GGA) is able to predict fairly reasonable {\it structural} energetics for Pu metal, with a prediction for the equilibrium lattice constant of $\delta$-Pu within $2.5\%$ of experiment\cite{review:SODERLIND:2019}. This is reasonably good and constitutes a significant improvement upon non-magnetic DFT. Of course, the predicted spin moments are nearly $5~\mu_B$ and anti-ferromagnetically ordered. The advantage of this approximation is in its computational expediency, 

Accordingly in this paper, we compare three different approaches to the calculation of formation energies and volumes of intrinsic defects in $\alpha$ and $\delta$-Pu: (i) collinear SP-GGA within the generalized gradient approximation at the theoretical equilibrium density, (ii) SP-GGA with an applied hydrostatic tension to approximate the experimental bulk density, and (iii) non-collinear SP+SO+OP-GGA at the theoretical density. 
We consider SP+SO+OP-GGA  to be the benchmark of the three approximations considered here.

While inclusion of SO and OP greatly improves accuracy, they dramatically increase the computational cost (still far less than a corresponding DMFT calculation).
Therefore, we characterize the corrections obtained from SO and OP and explore less expensive ways to obtain reliable point defect properties.
We argue that  a rough correction to the bulk SP-GGA overbinding  is simply to add a uniform external stress (in effect a $P V$ correction to the functional), 
thereby dilating the system to the approximate experimental density.
We find that this approximation works well so long as the lattice defects do not cause changes in the local magnetic ordering. 

In Section~\ref{sect:results}, we examine point defects in $\alpha$- and $\delta$-Pu, and compare formation energies and volumes to those in regular transition metals. The most important finding is the stark contrast in defect properties in $\alpha$- and $\delta$-Pu. Our results show opposite to $\alpha$-Pu as well as most close-packed metals: (i) the equilibrium concentration of vacancies in $\delta$-Pu is smaller than that of self-interstitials, and (ii) the magnitude of the lattice strain caused by the vacancies is nearly equal to that of self-interstitials. As a result, we estimate in Sect.~\ref{sect:discussion}, a small void swelling bias in $\delta$-Pu according to the conventional theory. In contrast, $\alpha$-Pu is argued to exhibit a swelling bias comparable to normal transition metals. 

At first glance, $\delta$-Pu appears characteristically as the outlier with unusual point defect properties, while they seem normal in $\alpha$-Pu. However, $\alpha$-Pu is a low-symmetry phase with  many inequivalent sites, not unlike metallic glasses. Why then do its point defects have relaxation volumes so similar to e.g. close-packed copper? It turns out that point defects in Pu crystals induce changes in sizes of the spin/orbital moments of nearby lattice sites. The magnitudes of these induced moments strongly correlate with the defect formation volumes. Hence, the large formation volumes in $\alpha$-Pu stem from defect-induced increases in sizes of the spin/orbital moments of the neighboring lattice sites, while introduction of point defects in $\delta$-Pu reduces spin-polarization in their vicinity, which in turn strengthens the effective interatomic bonding and causes vanishing defect formation volumes. Quantitative analyses of the effect of point defects on the local electronic structure of $\alpha$- and $\delta$-Pu phases are conducted throughout Sect.~\ref{sect:results}, using a novel thermodynamic variable, the so-called {\it {formation spin moment}}, introduced in Sect.~\ref{sect:thermo}.

We should emphasize in particular the value of this new analysis. The newly introduced formation spin measure highlights the changes to the local electronic structure (at this level of DFT) that correlate with defect-induced lattice strains. The spatial distribution of the latter can be quantified and categorized as elastic or beyond by studying changes in the Voronoi volumes of the nearby atoms. Note that changes in local electronic structure offer low energy pathways to altering the effective atomic bonding and in turn the defect formation volumes. Simplifying the situation by introducing a hypothetical electronic excitation cost, $\Delta E$, an associated volume change $\Delta V$, and a local pressure, $P$, such accommodations are favored if  
$-P\Delta V/\Delta E >1$. The accommodation is not symmetric - local moments in $\delta$-Pu are near the maximum achievable, while they are among the lowest possible in $\alpha$-Pu. This asymmetry can account for the strikingly different defect properties in the two phases. There are likely to be additional electronic degrees of freedom that contribute, e.g. changes in the local spin order from antiferromagnetic to ferromagnetic can affect the local atomic volumes and thereby defect-induced lattice strains. Nevertheless, the two Pu phases are in some sense at opposite extremes, and that basic difference is what underpins their distinct point defect properties. This suggests that successive levels of DFT (e.g. static SP, SP+SO+OP, up to and including full quantum fluctuations) should broadly agree as to these defect properties for the two phases.

To reiterate, our {\it ab initio} results suggest that the classic void-swelling bias is far larger in $\alpha$-Pu  than in $\delta$-Pu \cite{ALLEN:2015a,ALLEN:2015b}, thus swelling is more likely in the $\alpha$-Pu phase. This illustrates a broader lesson: different crystal phases of the same material can behave very different under irradiation.
However, conventional void swelling theory may not be reliable for predicting plutonium aging over long times. In $\delta$-Pu, defect interactions likely have magnetic contributions with anomalous dependences on temperature and internal stress-field, resulting from significant entropic contributions due to magnetic fluctuations. A few examples of low-lying magnetic excitations of the vacancy in $\delta$-Pu are discussed in Sect.~\ref{sec:vacancydelta}. In $\alpha$-Pu, the low symmetry and the quasi-two-dimensional diffusion of point defects in this system pose challenges to the classic mean-field theory of void swelling. The strongly anisotropic diffusion of point defects in $\alpha$-Pu is discussed in Sect.~\ref{sec:vacancyalpha}, where the distribution of defect formation energies among the eight different crystallographic sites are studied. The lowest-energy sites are arranged in two-dimensional layers, separated by energetically unfavorable regions. This anisotropy/inhomogeneity could also have implications for vacancy-mediated transport of substitutional He and Ga and thus bubble formation and phase metastability.

Nevertheless, in the short-time transient regime, our ab-initio predictions of the density changes of $\alpha$-Pu being much larger than $\delta$-Pu should be reliable and observable in experiments.

The paper is organized as follows. In \sect{sect:methodology}, we outline the details of our calculations, as well as a variational formulation of the orbital polarization method within non-collinear spin-density functional theory, implemented in the framework of the projector augmented wave (PAW) method. It allows for accurate and efficient computations of atomic forces and stresses, with application to structural relaxations induced by point defects in Pu lattices. In \sect{sect:thermo}, we review defect thermodynamics and introduce novel thermodynamic variables for analyzing defect-induced electronic transitions in Pu phases. In \sect{sect:results}, we present extensive calculations and analyses of $\alpha$- and $\delta$-Pu phases as well as a thorough study of point defect structures and energies in these phases. In \sect{sect:discussion} the classic void swelling theory is recapitulated, whereupon the swelling biases for the two Pu phases are discussed. 

\section{Methodology}
\label{sect:methodology}

The aim of this paper is to study from first principles the structures and energies of point defects in $\alpha$- and $\delta$-phases of Pu. The present work relies on static mean-field treatment of magnetism in Pu through two exchange-correlation functionals: SP-GGA and SP+SO+OP-GGA. The latter requires an order-of-magnitude more computational resources than the former. In this study, SP-GGA calculations are carried out for point defects in both $\alpha$ and $\delta$-Pu phases, while SP+SO+OP-GGA is only applied to point defects in $\delta$-Pu. The PBE\cite{PBE} parametrization of the GGA exchange-correlation functional is used throughout this work.

The defect calculations presented in the following are performed using periodic supercells containing 128 atoms for $\alpha$-Pu and 108 atoms for $\delta$-Pu. For accurate representation of the Fermi surfaces, the Brillouin zones of the 108/128-atom supercells are sampled with 27/8 k-points. The Kohn-Sham wave functions are represented in a projector-augmented wave (PAW) basis, as implemented in the Vienna Ab-initio Simulation Package (VASP)\cite{KresseFurth:1996}. A planewave cutoff of 450 eV is employed. The relativistic effects are taken into account by addition of a spin-orbit (SO) coupling term to the Hamiltonian. Structural relaxations induced by point defects are extensively studied following atomic forces derived within the PAW scheme.

The treatments of spin-orbit coupling and orbital polarization in this paper are unconventional and need explanation.  First, let us discuss our treatment of spin-orbit coupling in Pu. A comparative study of PAW and all-electron calculations of equilibrium structural properties of the light actinide metals (Th-Pu)\cite{SADIGH:2019} found that while for collinear spin-polarized calculations, the different methods are in good agreement, the results differ when SO coupling is included. It was found  that the incompleteness of the scalar-relativistic PAW basis used in VASP, i.e. the absence of the $p_{1/2}$-orbital, is responsible for this failure. This situation can be mitigated in actindes by discarding the coupling of the spin degrees of freedom to the $p$-angular momentum \cite{SADIGH:2019}. In all calculations presented below, we follow this strategy for incorporation of the SO coupling.

Next, we discuss the implementation of the LDA+U method and the orbital polarization technique within the PAW scheme in VASP. As a result of the relatively strong correlations in Pu metal, atomic orbital degeneracies in the $f$-electron subspace may be broken,  This cannot be accounted for within standard spin-density functional theory derived from the homogeneous electron gas. However, DFT can be generalized to treat orbital ordering in strongly correlated systems by incorporating on-site screened Coloumb interactions. The LDA+U formalism\cite{anis_zaanen_andersen:1991,gunnarsson:1991,lichtenstein:1995,Terakura:1998} is the simplest realization of this idea. It is a static mean-field approach that leads to addition of renormalized Hartree-Fock-like terms to the DFT Hamiltonian.
The two main disadvantages of the LDA+U theory are (i) the results are sensitive to parametrization of the screened local Coloumb interactions, and (ii) it is difficult to exactly account for double-counting of interactions. 

For treatment of orbital magnetism in itinerant systems, Brooks and coworkers\cite{ERIKSSON:1990} proposed a simplified theory, which in the presence of spin-orbit coupling accounts for orbital ordering originating from atomic Hund's second rule coupling, They proposed to augment the standard spin-density functional theory (SP-DFT) total-energy functional with a quadratic term, which in the context of collinear magnetism and $f$-electron orbital ordering can be written as 
\begin{equation}
  E_{\text{SP+SO+OP}} = E_{\text{SP+SO}} - \frac{1}{2} \sum_I E_I^{3} \left[\expval{L_{I,z}^f}\right]^2,
  \label{eq:OPB}
\end{equation}
where $\expval{\vec{L}_I^f}$ is the expectation value of the total orbital moment of the $f$-electrons at site $I$. It can be formally written as 
\begin{equation}
  \label{eq:Lhat}
  \expval{\hat{\vec{L}}^f} = \text{Tr}(\hat{\vec{L}}\hat{n}^f),
\end{equation}
where $\hat{\vec{L}}$ is the angular-momentum operator and $\hat{n}^f$   is the site-diagonal single-particle density matrix, expressed in the basis $\ket{3,m,s}$ of atomic $f$-orbitals
\begin{eqnarray}
  &\hat{n}^f& = \sum_{m,s,m',s'} n_{ms,m's'}^f~ \ket{3,m',s'}\bra{3,m,s}~~\\
  &n^f_{ms,m's'}& = <3,m',s'|n(r,r')|3,m,s>.  \nonumber
\end{eqnarray}

Additionally, $E^3_I$ in Eq.~\ref{eq:OPB} is the so-called Racah parameter, which in the case of $f$-orbitals can be expressed as a linear combination of the $F_2$, $F_4$ and $F_6$ Slater integrals of the $f$-orbitals within the atomic sphere around the $I$th nucleus. Equation~\ref{eq:OPB} was originally designed for collinear magnetic systems treated within the second variation method \cite{ANDERSEN:1975}. In this formulation $E^3_I$ can be calculated self-consistently, but that leads to complications in a fully variational treatment for calculations of the atomic forces, see below. 

The advantage of the SP+SO+OP-DFT formalism is that the Racah parameter $E^3$ is quite insensitive to environment and does not change significantly between the atom and the solid. Hence it can be calculated once and treated as a constant. In this way, a major drawback of the LDA+U formalism is resolved. However, the problem with this amendment of the SP-DFT functional is that in nearly all systems, it is too weak to have any significant effect on structural energies. It has received most attention for its application to ab-initio calculations of magneto-crystalline anisotropies\cite{trygg}. In a recent study of the structural enegies of light actinides\cite{SADIGH:2019}, it was found that the addition of OP to the SP+SO-GGA functional has a small effect on the structural energies of all the light actinide elements except for Pu. In this system, addition of OP remarkably improves upon the SP+SO-GGA predictions of the equilibrium volumes of both the $\alpha$-Pu as well as the $\delta$-Pu phases, while yielding correct energy ordering of the phases. In the same paper, a comparative study with the popular GGA+U scheme devised by Dudarev and coworkers\cite{dudarev:ldau} was conducted for the $\alpha$- and the $\delta$-Pu phases. It led to the conclusion that for the $\alpha$-Pu phase to remain the ground state phase of Pu metal, the $U$ parameter may not exceed 0.2~eV, whereupon it underestimates the equilibrium volumes of the $\alpha$- and the $\delta$-Pu phases by 3-4\%, compared to 1\% within the SP+SO+OP-GGA formalism.

The SP+SO+OP functional, Eq.~\ref{eq:OPB} can be easily implemented within the PAW formalism following Bengone {\it et al.}\cite{Bengone:2000}.  However, in order for the Hellmann-Feynman theorem to be applicable to derivation of the atomic forces in the presence of SO and OP, a fully variational implementation using spinor wave functions and  non-collinear magnetism\cite{KresseHobbs:2000} is required. Furthermore the Racah parameter $E^3$ must be treated as a constant. Such a formulation has been developed and tested in a previous paper\cite{SADIGH:2019}. In this scheme, the quadratic term in Eq.~\ref{eq:OPB} is generalized for arbitrary spin orientations, leading to the following total-energy expression:
\begin{align}
\label{eq:1}
E_\text{SP+SO+OP} &= E_{\text{SP+SO}}- \frac{1}{2} \sum_I E^3 \frac{\left(\expval{\hat{\vec{L}}^f_{I}}\cdot\left<\hat{\vec{S}}^f_{I}\right>\right)^2}  
{\left<\hat{\vec{S}}^f_{I}\right>\cdot\left<\hat{\vec{S}}^f_{I}\right>},
\end{align}
where $\expval{\hat{\vec{L}}^f_I}$ and $\expval{\hat{\vec{S}}^f_I}$ are site-projected spin and orbital moments calculated following Eq.~\ref{eq:Lhat}.  Details of implementation within the PAW method can be found in Ref.~\onlinecite{SADIGH:2019}. 

In this paper, we use the variational total-energy functional Eq.~\ref{eq:1} implemented within the PAW scheme in the VASP code for SP+SO+OP-GGA calculations of Pu. The calculations are parametrized by the value $E^3=0.0621$ eV, obtained as an average of spin-decomposed Slater integrals of site-projected $f$-electron wave functions in $\delta$-Pu at equilibrium lattice constant, using the all-electron Full-Potential Linear Muffin-Tin Orbital (FPLMTO) technique \cite{WILLS:2010}. This parameter is kept fixed throughout the calculations in this paper. 

Below, we further validate the PAW approach to the SP+SO+OP-GGA functional, by comparing its predictions for properties of $\delta$-Pu with all-electron calculations within the FPLMTO scheme\cite{WILLS:2010}, using the second variation method \cite{ANDERSEN:1975} to SO and OP. The details of these calculations are quite similar to those applied previously to investigations of the Pu phase-diagram\cite{SODERLIND:2004}. The present FPLMTO implementation does not make any assumptions beyond the GGA. Basis functions,
electron densities, and potentials are calculated without any geometrical approximation, and these are
expanded in spherical harmonics inside non-overlapping (muffin-tin, MT)
spheres surrounding each atom and in Fourier series in the region between these muffin-tin spheres.
One has to define an MT sphere with a radius, s$_{MT}$, and here it is chosen so that s$_{MT}$/s$_{WS}$ $\sim$~ 0.8, where
s$_{WS}$ is the Wigner-Seitz (atomic sphere) radius. The radial parts of the basis functions inside the MT
spheres are calculated from a wave equation for the L = 0 component of the potential that includes all
relativistic corrections including spin-orbit coupling for $d$ and $f$ states, but not for the $p$ states, which has been previously shown to be an appropriate and accurate procedure \cite{SADIGH:2019}.

\section{Point defect thermodynamics}
\label{sect:thermo}

\subsection{Formation enthalpy and volume}

The most basic thermodynamic property of a point defect in a crystal at pressure $P$ is its Gibbs free-energy of formation $\Delta G_F(P,T)$. It is the free energy change of the crystalline solid per every point defect introduced in it. The equilibrium concentration of point defects at finite temperatures can thus be easily derived to be
\begin{equation}
  c_{eq}(P,T) \propto \exp(-\frac{\Delta G_F(P,T)}{k_BT}),
\end{equation}
with the Gibbs free energy defined in terms of formation enthalpy $\Delta H_F$ and formation entropy $\Delta S_F$ as
\begin{equation}
  \Delta G_F(P,T) = \Delta H_F(P,T) - T\Delta S(P,T).
  \label{eq:GG}
\end{equation}

In what follows, we will be deriving expressions for mechanical properties of point defects in lattices at uniform equilibrium temperature. Hence, temperature only appears as a parameter in the equations. In this paper, we are only concerned with ab initio ground state calculations. Generalization to finite temperatures is straightforward, but here we specialize to $T=0$ and drop the formation entropy term in Eq.~\ref{eq:GG}.

The ab initio calculations of point defect properties in $\alpha$- and $\delta$-Pu presented below are performed using the standard periodic supercell technique. Hence, consider an $N$-atom supercell of the perfect crystal lattice at pressure $P$, with supercell volume $V_L(P)$ and enthalpy $H_L(P)$. A vacancy (interstitial) can be introduced by removing (adding) an atom from (to) this system. Denote by $N_D$ the number of atoms in the defect supercell with $N = N_D \pm 1$ (vacancy and interstitial, respectively). Similarly, denote by $V_D(P)$ the defect supercell volume and by $H_D(P)$ its enthalpy. The formation enthalpy of the point defect can be written as 
\begin{eqnarray}
\Delta H_F(P) &=& H_D(P) - N_D \frac{H_L(P)}{N}.\\
 &=& H_D(P) - H_L(P) + (N - N_D) \frac{H_L(P)}{N} \label{eq:hf_2}.
\end{eqnarray}

The significance of reordering of terms to obtain Eq.~\ref{eq:hf_2} is that if a generalized specific enthalpy $\tilde{H}(P,c)$ as a function of both pressure $P$ and defect concentration $c$ is defined such that $\tilde{H}(P,0) = H_L(P)/N$ and $\tilde{H}(P,1/N) = H_D(P)/N$, then in the limit $N\rightarrow \infty$, we have   
\begin{equation}
  \Delta H_F(P) = \frac{\partial\tilde{H}(P,0)}{\partial c} \pm \tilde{H}(P,0).
\end{equation}
for vacancy and interstitial, respectively. 

This transformation allows for a more transparent definition of the formation enthalpy and thereby a simpler derivation of other thermomechanical properties of point defects. As an example, let us consider the first pressure-derivative of the formation enthalpy. Using the thermodynamic relation $\frac{\partial H}{\partial P} = V$, we obtain an expression for the formation volume
\begin{equation}
  \Delta V_F(P) = \frac{\partial^2 \tilde{H}}{\partial P \partial c} = \frac{\partial\tilde{V}(P,0)}{\partial c} \pm \tilde{V}(P,0),
  \label{eq:relV}
\end{equation}
where $\tilde{V}(P,c)$ is a generalized specific volume that is defined in such a way as $\tilde{V}(P,0) = V_L(P)/N$ and $\tilde{V}(P,1/N) = V_D(P)/N$. Of course, in practice, the partial derivative is replaced by the finite difference formula
\begin{equation}
  \Delta V_R(P) \equiv \frac{\partial\tilde{V}(P,0)}{\partial c} = V_D(P) - V_L(P).
  \label{eq:relV1}
\end{equation}

Above we have defined the relaxation volume $\Delta V_R(P)$, which is obviously related to the formation volume $\Delta V_F$, but is more relevant to materials mechanics and is not directly observable in macroscopic experiments. The formation volume on the other hand can be measured experimentally via e.g. dilatometry. It quantifies the change in size of a solid when point defects are incorporated in it through damage accumulation. Of course, point-defect induced volume changes are most relevant at the earliest stages of damage before any substantial defect clustering, such as e.g. void nucleation has occurred. 

\subsection{Elasticity of point defects}
\label{sec:compress}

The formation volume provides a measure for how well a crystal lattice can macroscopically accommodate point defects. When $\Delta V_F \approx 0$, the host crystal is maximally accommodative, and thus does not show any macroscopic size change upon damage accumulation. However, the macroscopic accommodation is accompanied by internal microscopic strain fields in the solid. This is because, e.g. a vacancy/interstitial with zero formation volume, induces inward/outward displacements of the nearest neighbor shell, which propagates through the lattice as a long-range compressive/tensile strain field. This long-range displacement field is responsible for the elastic interaction with other lattice defects, e.g. other point defects, dislocations and grain boundaries. The strength of the strain field induced by a point defect is quantified by the relaxation volume $\Delta V_R$. The latter is large in magnitude whenever $|\Delta V_F|$ is small. 

In order to rigorously model a point defect or impurity in a lattice as an elastic inclusion,  its effective size and elastic properties must be able to be uniquely specified from atomistic first principles. As an example of the effective elastic properties of the point defect, we discuss here the effective point-defect compressibility, which we define as
\begin{equation}
  \Delta \beta = -\frac{1}{\Omega_L} \frac{\partial \Delta V_R(P)}{\partial P} = -\frac{\partial \Delta V_R/\Omega_L}{\partial P} + \frac{\Delta V_R}{\Omega_L}~\frac{1}{K_L},
  \label{eq:dbeta}
\end{equation}
where $\Omega_L = \tilde{V}(P,0)$ is the atomic volume, and $K_L$ is the bulk modulus of the perfect crystal. The right-hand side expands $\Delta \beta$  into two terms. This is useful, since the relaxation volume is most conveniently expressed in units of atomic volume. In practice, the two contributions to the compressibility can be calculated in finite periodic supercells as follows: 
\begin{eqnarray}
  \label{eq:beta}
  \Delta \beta &=& \Delta \beta_0 + \Delta \beta_L, \\
  \label{eq:beta0}
  \Delta \beta_0 &=& \left(\frac{\Delta V_R}{\Omega_L} + N\right) \left(\frac{1}{K_D} - \frac{1}{K_L}\right), \\
  \label{eq:betaL}
  \Delta \beta_L &=& \frac{\Delta V_R}{\Omega_L}~\frac{1}{K_L},
\end{eqnarray}
where $N$ is the number of lattice sites in defect supercells, i.e. the inverse of the defect concentration, and  $K_D$ is the bulk modulus of the supercell containing one point defect. It is worth noting that $\beta_0$ represents the contribution to the compressibility of point defects derived from their effect on the lattice stiffness, i.e. $\Delta \beta_0$ vanishes whenever $K_D = K_L$, while $\Delta \beta_L$ remains non-zero so long as $\Delta V_R \neq 0$.

\subsection{Formation spin moments}

The unusual electronic structure of Pu metal is responsible for its polymorphic phase diagram, as well as a slew of other anomalous thermophysical properties. In particular, the $\alpha$- and the $\delta$-Pu phases distinguish themselves from each other by having low crystal symmetry/high density, and high crystal symmetry/low density, respectively. The structural chemistry within our DFT approach is largely determined by a competition of two symmetry-breaking mechanisms enabled because the delocalized or itinerant 5$f$-electron density of states is (i) very narrow and (ii) positioned close to the Fermi level. On the one hand, degenerate energy states can be lifted due to a Jahn-Teller or Peierls like crystal distortion that lowers the total energy for lower-symmetry crystal structures \cite{SODERLIND:1995}. On the other hand, the high-energy degenerate states can spin polarize and form magnetic moments. The exchange interaction shifts states away from the Fermi level to a lower energy while occupying orbitals with greater anti-bonding characteristics leading to volume expansion.\cite{SODERLIND:2008} The former  mechanism favors the $\alpha$ phase while the latter favors the $\delta$ phase. In the real material both these effects, with contributions from  spin-orbit coupling and orbital polarization, favor one phase over another in a delicate shifting balance (plutonium has six ambient phases with varying degrees of crystal symmetry, atomic density, and magnetic moments).

This competition between crystal distortion and magnetic-moment formation can be quantified in the different Pu phases as well as at different atomic sites.\cite{SODERLIND:2004} The magnitudes of site-projected spin moments, calculated within spin/orbital-polarized DFT, offers one such measure. Of course, the static spin/orbital-polarized DFT potential is derived for weakly-interacting itinerant electrons, and thus fails to capture any physical properties that may involve multi-configuration correlated electronic states.  Nevertheless, it has been quite successful in describing the structural chemistry of plutonium \cite{SODERLIND:2004}, by allowing for formation of static spin moments. 

Accordingly here, we define the average magnitude of spin moments per atom in a system with $N$ Pu sites
\begin{equation}
  \Sigma = \frac{1}{N}\sum_{i=1}^{N} \left|<\vec{\hat{S}}_i>\right|,
  \label{eq:avmom}
\end{equation}
where $<\vec{S}_i>$ are site-projected spin moments at each site $i$, obtained from projection of the spin-density matrix onto the respective muffin-tin sphere. $S_i$ are scalar quantities for collinear SP-GGA, and they become 3-vectors for non-collinear SP+SO-GGA as well as SP+SO+OP-GGA schemes. It can also be defined within more elaborate and realistic approximations beyond DFT. 

We will discuss in the next section the values of $\Sigma$ in $\alpha$- and $\delta$-Pu within different levels of theory. It turns out that it is e.g. more than twice larger in $\delta$-Pu ($\Sigma^{\delta} = 4.4$) than in $\alpha$-Pu ($\Sigma^{\alpha}=2.1$), when calculated within the SP+SO+OP-GGA. This then naturally explains why the $\delta$ phase has far larger atomic volume
than the $\alpha$ phase. Significantly, it also implies that $\alpha$-Pu is far from a simple metal with meaningful $f$-electron correlation, although less than $\delta$-Pu.

Later in this paper we show the power of $\Sigma$ to provide understanding of the electronic structure as well as the thermodynamic properties of point defects in $\alpha$- and $\delta$-Pu. For this purpose, we consider the same $N$-atom supercell as above representing a perfect crystal lattice of Pu metal at pressure $P$, with average moment per site $\Sigma_L(P)$. Introducing a point defect in this supercell results in $N_D$ Pu atoms and an average moment per site of $\Sigma_D(P)$. The formation moment of the point defect $\Delta \Sigma_F$ can now be written as  
\begin{equation}
  \Delta \Sigma_F = N_D (\Sigma_D(P) - \Sigma_L(P)).
  \label{eq:fmom}
\end{equation}
It will be shown below that $\Delta \Sigma_F$ can be a powerful measure of the effect of point defects on $f$-electron hybridization in different phases of Pu. Furthermore, it will become clear that this contribution is of crucial importance to the point defect properties in both $\alpha$- and $\delta$-Pu, and consequently to their resistance to self-irradiation-induced swelling.

\section{Results}
\label{sect:results}

\subsection{Pure Pu}
\label{sec:purePu}

 Before embarking on a discussion of point defects in Pu, we first compare the situation for the two bulk phases of interest.
By now, there is considerable experience with first principles results on these phases \cite{SODERLIND:2001,SADIGH:ALPHA:2003,SODERLIND:2004,SADIGH:2019}. 
They are quite close in energy at ambient pressure despite very different crystal structure and density.

Nonmagnetic (NM) scalar-relativistic DFT calculations obtain an equilibrium volume of  ~16.5 \AA$^3$ for the $\delta$-Pu phase, which is a drastic underestimation of the experimental value $\sim 25$~\AA $^3$. The situation is better for the $\alpha$-phase, for which NM-GGA predicts an equilibrium volume of ~17.6~\AA $^3$.  This still corresponds to an overbinding of $\sim10$\% as compared with the measured value at low temperatures $\sim 19.3$~\AA $^3$. 

The inclusion of spin polarization and magnetism in the SP-GGA calculations corrects to a large extent the equilibrium volume of the $\delta$-Pu phase (to ~23.0 \AA$^3$) and also brings the calculated $\alpha$-Pu volume to ~18.2 \AA$^3$, in closer agreement with experiments. 
However, very large spin moments ($\sim 5\mu_B$) are predicted to form on each Pu atom in $\delta$-Pu. Surprisingly, even in $\alpha$-Pu, 
several Pu sites have large spin moments $\sim 3.5 \mu_B$, and all Pu atoms possess some spin polarization. Hence, moment formation is ubiquitous in all phases of Pu within DFT. 

Incorporation of SOC generally reduces the calculated spin moments by a small amount, e.g. in $\delta$-Pu, they are reduced to $\sim 4.4 \mu_B$. Furthermore, SOC induces orbital moments opposite to the spin moments on each Pu site, which further reduces the total magnetic moment per Pu site in $\delta$-Pu to $\sim 2.3 \mu_B$.  

Finally, inclusion of OP in the Hamiltonian increases the magnitude of the orbital moment per Pu site and consequently reduces the total magnetization of each $\delta$-Pu atom to 0.74, in excellent agreement with recent DMFT calculations\cite{JANOSHEK:2015}. Furthermore, predicted densities are in good agreement with experiment at 24.7 and 19.25 \AA$^3$ for $\delta$ and $\alpha$, respectively.

In the following, we separately discuss calculations and measurements of some common physical properties of (pure) $\alpha$-Pu and $\delta$-Pu at  ambient conditions.  We include analyses at both theoretical and experimental ambient conditions for SP-GGA since the two limits differ greatly.

\begin{table*}[htbp]

\caption{Properties of pure $\delta$-Pu using VASP PAW:
SP denotes collinear spin polarized calculations with AF order; the -3 GPa external pressure approximates the experimental volume. SO refers to inclusion of spin-orbit coupling in a non-collinear spinor representation, while neglecting the contribution from the $p$-orbitals, and OP to the addition of orbital polarization correction.
}

\centering
\begin{tabular}{l c c c c c c c}
\hline\hline

Property                             & NM                  & SP             &  SP                                     & SP+SO                          \comnt{&  SP+SO (non-coll)  }            & SP+SO+OP            & SP+SO+OP          & experiment\\ 
                                          &                         & (0 GPa)             & (-3 GPa) \comnt{(-29.54kbar)} & \comnt{(~coll)}                 \comnt{&                                  }           & (001) polarization      & (110) polarization     &                    \\ [1ex]
 \hline
Eq. volume (\AA $^3$)      & 16.5                 & 23.0                 &  24.75                                       &  23.6                                \comnt{&                                  }     & 24.8       & 24.7                                    & 25.2\cite{PuHand}\\
Bulk modulus (GPA)          & 208                & 44.7                 &  36.0                                         & 45.0                                 \comnt{&                                 }       & 44.6       & 39.1                                     &  30-35 \cite{MOMENT76}\\
c/a                                      & 0.706              & 0.942               &  0.991                                        & 0.963                               \comnt{&                                  }    & 0.993    & 0.982                                  & 1.0  \\
spin mom. ($\mu_B$)         & -                     & 4.78                 &  4.95                                          & 4.39                                 \comnt{&                                  }    &  4.4                                           &  4.4                         &    -    \\ 
magnetic mom. ($\mu_B$) & -                     & -                       & -                                                & 2.34                  \comnt{&                                  }    & 0.78                                        & 0.74                       &  - \\[2ex]
\hline
\end{tabular}
\label{r-tab:1}
\end{table*}

\subsubsection{$\delta$-Pu}
\label{sec:puredeltaPu}

The properties of $\delta$-Pu phase as calculated within different levels of theory are listed in Table~\ref{r-tab:1} and are compared with experiments.  The drastic underestimation of the equilibrium volume by the NM-GGA theory, largely corrected by inclusion of spin and orbital polarization is clearly documented.

The  SP-GGA calculations are constrained to have collinear spins. This limitation is relaxed upon the inclusion of relativistic effects via non-collinear SO-coupling, but even then the local spin moments do not appreciably deviate from collinear AF ordering. This implies that this spin configuration is near a local minimum of the potential-energy landscape when generalized to non-collinear magnetism.

Since the SO-treatment implemented within the VASP code does not explicitly account for the $p_{1/2}$-states, a non-negligible error is introduced, which can be mostly eliminated by simply removing the spin-orbit ($L\cdot S$) matrix elements corresponding to the $L=1$ angular momentum.\cite{SADIGH:2019} A comparison between results obtained with and without the $p$-channel SO-coupling is shown in \tab{r-tab:1.1}. Hence, proper inclusion of SO expands the lattice by $\sim 3\%$, in better agreement with experiments, but still about $\sim 6\%$ too small.

\begin{table}{}
\caption{Properties of pure $\delta$-Pu calculated by FPLMTO. SO$_{all}$ denotes SO-coupling including all angular momentum channels, and SO$_{no p}$ refers to SO-coupling excluding the $p$-channel. 
 }
\centering
\begin{tabular}{c c c c}
\hline\hline
Method & Eq. volume  & Bulk modulus & c/a \\ 
         &  (\AA $^3$) & GPa                &     \\[1ex] 
\hline
SP+SO$_{all}$   & 22.9  & 40. & 0.92  \\
SP+SO$_{no~p}$  & 23.7 & 42 & 0.96   \\
SP+SO$_{all}$+OP& 24.6 & 45 & 0.992   \\
SP+SO$_{no~p}$+OP & 24.8 & 45 & 0.993 \\ [2ex]
\hline
\end{tabular}
\label{r-tab:1.1}
\end{table}

\begin{table}{}
\caption{Properties of pure $\alpha$-Pu calculated by PAW.}
\centering
\begin{tabular}{c c c c c c}
\hline\hline
Method & Eq. volume  & Bulk modulus & b/a & c/a & $\theta$ \\ 
         &  (\AA $^3$) & GPa        &     &   &     \\[1ex] 
\hline
NM &  17.6   & 187.5 & 1.85  & 0.754  & 102.2\\
SP  & 18.2  & 103.6 & 1.83  & 0.757 & 101.7\\
SP+SO  & 18.35 & 141.6 & 1.82 & 0.76 & 101.9   \\
SP+SO+OP  & 19.25 & 57.4 & 1.79 & 0.75 & 101.5   \\
experiment (0~K) & 19.5 & 70.9 \cite{alphaBulkMod} & 1.77 & 0.755 & 101.8 \\ [2ex]
\hline
\end{tabular}
\label{tab:alpha}
\end{table}

\begin{figure}
\includegraphics[trim= 0 220 300 0, width=1.00\linewidth]{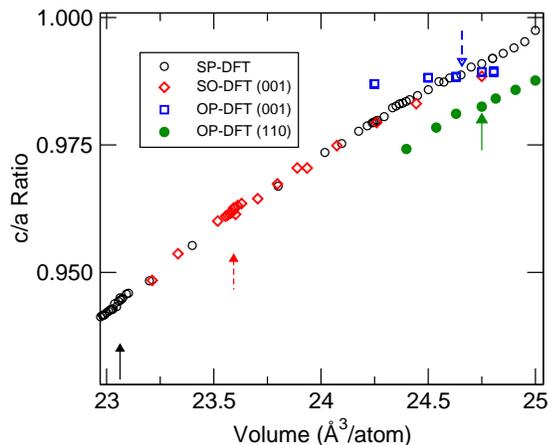}
\caption{Predicted c/a ratios versus volume for  bulk $\delta$-Pu in SP-GGA (open black circles), SO-GGA (open red diamonds), and OP-GGA (open blue squares for spin polarized on (001) and filled green circles for (110) polarization).  The arrows mark the theoretical equilibrium volume for the four cases (SP-GGA in solid black, SP+SO-GGA in dashed red, (110)-polarized SP+SO+OP-GGA in solid green, and (001)-polarized SP+SO+OP-GGA in downward-pointing, dashed blue).}
\label{fig:1}
\end{figure}

Following established literature, we represent in this paper the magnetic structure of the $\delta$-Pu phase by the collinear antiferromagnetic L10 order, known to be the lowest energy bulk magnetic order in the SP-GGA approximation. The layered structure breaks the cubic symmetry of $\delta$-Pu and results in a tetragonal distortion of the fcc lattice. Figure~\ref{fig:1} shows that the degree of tetragonality, as quantified by c/a ratio of the resulting face-centered tetragonal lattice, is strongly density dependent. The distortion is substantial ($\sim$ 6\%) at the SP-GGA equ  ilibrium volume of $23$~\AA $^3$, whence as atomic volume is increased, so does also the c/a ratio almost proportionally and approaches unity (cubic symmetry) near the experimental equilibrium volume of $25$~\AA$^3$. 

Spin-orbit coupling introduces additionally a spin quantization axis, which causes the energy to also depend on the direction of spin polarization; this is inherited by the OP approximation as well. This magnetic anisotropy is expected to have a small effect on structural chemistry. However, as shown in Fig.~\ref{fig:1}, it does noticeably change the volume dependence of the degree of tetragonality. It can be seen that the c/a ratios for (110)-polarization, calculated within both SP+SO-GGA and SP+SO+OP-GGa, are about 1\% shifted away from cubic symmetry compared to SP-GGA values for the same Pu density. However, the effect of the (001)-polarization on the c/a ratios is a bit more complicated. When calculated within the SP+SO-GGA approximation, this polarization induces quite similar c/a ratios to the SP-GGA calculations. In contrast, when the OP is included, the c/a ratios become nearly insensitive to the Pu density, and vary within a narrow interval (0.99,1). This emphasizes the role of orbital polarization in determining the structural energetics of Pu metal.

To understand the observed behavior in Fig.~\ref{fig:1}, one needs to note that spin-orbit coupling at a site with a finite spin moment induces an orbital polarization opposite in direction. Addition of explict OP in the Hamiltonian Eq.~\ref{eq:1}, increases the magnitude of the orbital polarization. Hence, when AF ordering is along (001) but spin/orbital polarization is along (110), the degree of symmetry breaking resulting from the AF symmetry is maximal and insensitive to the magnitude of $E^3$ in Eq.~\ref{eq:1}. In contrast when the spin/orbital polarization is  parallel to the AF stacking along the (001) direction, the density-dependence of the degree of tetragonality is reduced with increasing $E^3$. This implies a subtle but very real and important interplay between spin ordering and orbital polarization. At current level of theory, the static (001) spin/orbital polarization parallel to the AF stacking, in spite of its slightly higher energy compared to the perpendicular (110) polarization,  should be considered the better representative for the real spin-fluctuating system, as it stabilizes the crystal structure that comes closest to the high-symmetry cubic $\delta$-phase.


In contrast to the degree of tetragonality, the effect of SO and SO+OP corrections on equilibrium volume of the $\delta$-Pu phase is quite insensitive to the polarization direction, as seen in t~\ref{r-tab:1}. Addition of SO alone, only slightly expands the SP-GGA equilibrium volume towards the experiment. However, a much more substantial correction is obtained upon incorporation of OP, which approaches the equilibrium volume of $\delta$-Pu to within 2$\%$. It is easy to suppose that proper inclusion of magnetic fluctuations can restore full cubic symmetry.\cite{review:SODERLIND:2019}

\subsubsection{$\alpha$-Pu}
\label{sec:purealphaPu}

The $\alpha$-Pu phase is a monoclinic crystal of $P2_1/m$ symmetry \cite{alphaXtl} with a primitive cell containing 16 atoms. The space group includes inversion symmetry, which reduces the number of crystallographically inequivalent sites in the primitive cell to 8. Four parameters are necessary to specify the Bravais lattice: the length of the lattice vectors, denoted below by $a$, $b$, $c$, and the monoclinic angle $\theta$.
  
The experimental atomic volume of $\alpha$-Pu is 20.0 \AA$^3$ at 21 C\cite{lallement}. However, $\alpha$-Pu exhibits an unusually large thermal expansion at low temperatures, and it is estimated to have an equilibrium volume at $0~$K of about 19.5 \AA$^3$\cite{SADIGH:2019}. Nevertheless, the SP-GGA approximation still  overbinds this phase by about $\sim 7\%$. The equilibrium properties of $\alpha$-Pu, as calculated within different levels of theory compared to experiments are listed in Tab.~\ref{tab:alpha}. It is observed that the addition of OP expands the equilibrium volume to within $\sim 1 \%$ of the experimental value.

\begin{figure}
\caption{An image of $\alpha$-Pu structure with its eight inequivalent sites.}
  \includegraphics[clip,trim=2.cm 2.5cm 2.cm 4.cm, width=0.85\columnwidth]{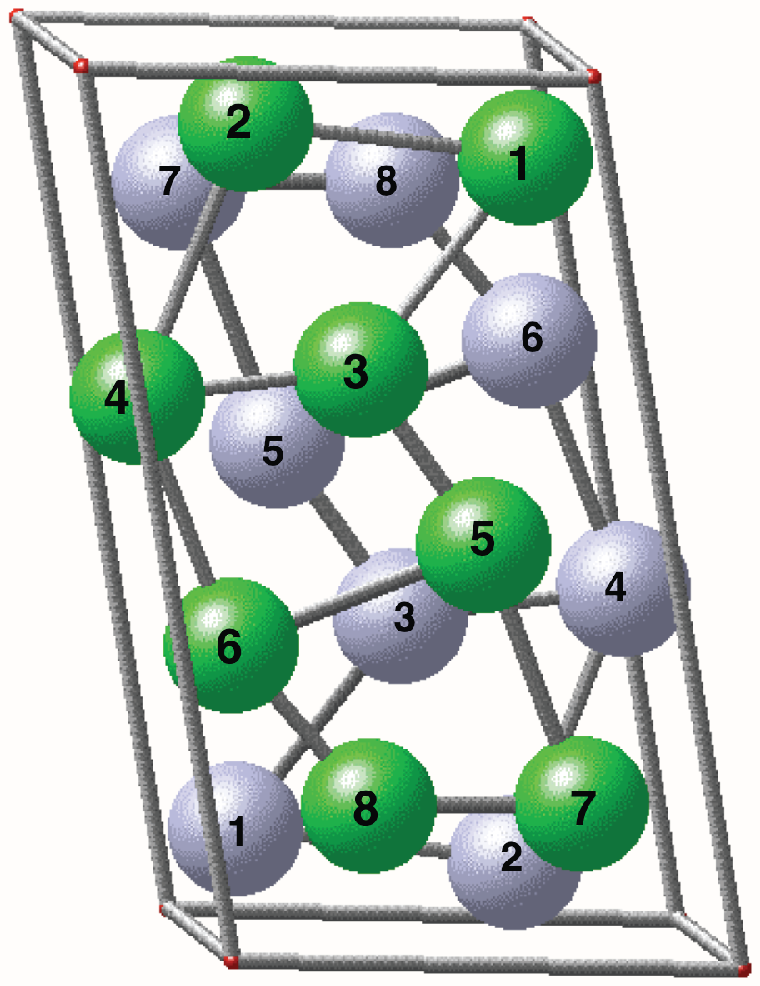} 
\label{FigEight}
\end{figure}

An important property of the $\alpha$-Pu phase, relevant to its alloying as well as lattice defect properties,\cite{GaStable:SADIGH:2005} is the significant difference in atomic volume and electronic properties of the eight inequivalent sites of this lattice. Table~\ref{tab:A_vor_dev} details the variation in the sizes of the Voronoi volumes of these eight sites, which are enumerated in accord with past literature.\cite{lallement} It is found that the smallest site (1) is nearly 20$\%$ smaller than the largest site (8). It is noteworthy that the local electronic structure does also vary strongly between the different sites. Table~\ref{tab:f_dev} lists the magnitudes of the local spin, as well as the magnetic moment at each of the eight sites. The dramatic difference between sites 1 and 8 is evident. The site-projected electronic densities-of-states for these sites have also been shown to be very different.\cite{review:SODERLIND:2019} While site 1 has the highest density (smallest Voronoi volume) and minimal spin polarization, site 8 is $\delta$-like with low density and large local spin moment. This heterogeneity plays an important role for the energies and volumes of point defects introduced into this phase.  

\begin{table}{}
  \caption{The table lists the percentage deviations of Voronoi volumes of all inequivalent sites of the $\alpha$-Pu phase from the average atomic volume. Comparison is made between room temperature experimental structure and equilibrium structures obtained from different levels of zero-temperature theory with static magnetism.}
  \centering
\begin{tabular}{c |c c c c }
\hline
Site       &      Experiment (room temp.)   &  SP         & SP+SO       &  SP+SO+OP         \\ [1ex] 
\hline
  1    &    -7.21$\%$                  & -6.35$\%$                     & -5.98$\%$                    & -7.80$\%$  \\
  2    &    -0.61$\%$                  & -1.49$\%$                     & -1.23$\%$                    & -0.87$\%$  \\
  3    &    -2.25$\%$                  & -2.13$\%$                     & -1.79$\%$                    & -1.89$\%$  \\
  4    &    -1.87$\%$                  & -3.47$\%$                     & -3.34$\%$                    & -3.33$\%$  \\
  5    &    -0.66$\%$                  & -1.83$\%$                     & -1.76$\%$                    & -0.67$\%$  \\
  6    &     0.67$\%$                  &  2.19$\%$                     &  1.66$\%$                    &  1.68$\%$  \\
  7    &     0.32$\%$                  &  1.92$\%$                     &  1.53$\%$                    &  0.99$\%$  \\
  8    &    11.63$\%$                  & 11.16$\%$                     & 10.91$\%$                    & 11.89$\%$  \\[1ex]
\hline

\end{tabular}
\label{tab:A_vor_dev}
\end{table}

\begin{table}{}
  \caption{The table lists the magnitudes of the local spin moments ($|\overline{S}_i|$), as well as the local magnetic moments ($|\overline{S}_i+\overline{L}_i|$ at each inequivalent site of the $\alpha-Pu$ phase, with $i$-index enumerating the 8 inequivalent sites. The calculations have been performed within the SP+SO and SP+SO+OP approximations at the equilibrium structure and density of $\alpha$-Pu within SP+SO+OP. }  
  \centering
\begin{tabular}{c |c c  | c c }
\hline
Site&\multicolumn{2}{c|}{SP+SO}&\multicolumn{2}{c}{SP+SO+OP}\\
\hline
i      &    $|\overline{S}_i|$         & $|\overline{S}_i+\overline{L}_i|$        &  $|\overline{S}_i|$         & $|\overline{S}_i+\overline{L}_i|$      \\
       &                 ($\mu_B$)                  &  ($\mu_B$)                              &                      ($\mu_B$)          &    ($\mu_B$)                           \\   [1ex] 
\hline
  1                   & 0.47                     & 0.35               & 0.55      & 0.14  \\
  2                   & 1.3                     & 0.73                & 2.0      & 0.18  \\
  3                   & 1.3                     & 0.7                 & 2.0     & 0.19  \\
  4                   & 1.4                     & 0.76                & 1.9      & 0.31  \\
  5                   & 1.3                     & 0.59                & 2.1      & 0.41  \\
  6                   &  1.9                     &  0.94              &  2.5       &  0.31  \\
  7                   &  2.3                     &  1.2               &  2.6      &  0.03  \\
  8                   & 3.5                     & 1.6                & 3.4      & 0.03  \\[1ex]
\hline

\end{tabular}
\label{tab:f_dev}
\end{table}

In summary, in contrast to the wide-spread expectation that ordinary non-magnetic DFT should be quite reasonable for describing the energetics of $\alpha$-Pu, it was found many years ago that quite sizeable spin and orbital moments form on several atoms in $\alpha$-Pu \cite{SADIGH:ALPHA:2003,GaStable:SADIGH:2005}.  Accordingly, quite large errors arise in predictions of the equilibrium volume of the $\alpha$-Pu phase, when neglecting spin and orbital polarization as well as spin-orbit coupling.

While the SO and OP approximations allow for non-collinear spins, we find that the lowest energy collinear configurations constitute local potential-energy minima. Our search for the lowest-energy collinear magnetic configuration in $\alpha$-Pu has led us to the ferrimagnetic order, with sites 1, 3, 4, 7 antiparallel to 2, 5, 6, and 8, and the magnitudes of the local spin and orbital moments listed in Tab.~\ref{r-tab:f_dev}. Because of the complexity of the $\alpha$-Pu structure and the diversity of possible point defect configurations in this phase,  we have limited the scope of this work to comprehensive point defect calculations in 128-atom $\alpha$-Pu supercells using the SP-GGA approximation only. Encouraged by the reasonable success of SP-GGA for the swelling parameters in $\delta$-Pu, we perform these calculations at negative stress of -5 GPa, corresponding to an equilibrium volume near the experimental value. In this way, we correct for the overbinding of this approximation,

\subsection{Point defects in Pu lattices}
\label{sec:defects}

In this section, we explore the energetics and the structures of the intrinsic point defects in the $\delta$- and $\alpha$-Pu phases.  
The quasi-cubic high-symmetry of the $\delta$-phase greatly simplifies the search for favorable defect geometries, 
while monoclinic $\alpha$-Pu offers a multitude of crystallographically-distinct defect sites each with low-symmetry local environments to relax. 
Accordingly, the defect study for the $\delta$-phase is more comprehensive, including for the first time a limited survey of changes in local magnetic order.
As with the bulk calculations described above, we explore different levels of DFT approximation to examine defect properties in $\delta$-Pu, including SP-GGA at the theoretical equilibrium, strained-SP-GGA under hydrostatic stress to approach the experimental density, and SP+SO+OP-GGA.

When strained-SP-GGA and SP+SO+OP-GGA calculations of point defects in $\delta$-Pu are initialized with the same magnetic order, their formation energies and volumes are found to be in reasonable quantitative agreement notwithstanding subtle differences in predicted defect structures. However, a sampling of magnetic structures in $\delta$-Pu reveals further complexity that will be discussed in detail in the next subsection,

Because the (inexpensive) strained-SP-GGA approximation seems adequate for a first look at swelling behavior in $\delta$-Pu, we use only this method to study intrinsic point defects in $\alpha$-Pu.
No comparison with SP+SO+OP-GGA is made for defects in $\alpha$-Pu at this time, and more detailed studies are left for future work. Instead, we focus below on exploring the energetics and structures of the multitude of point defect configurations that are made possible by the low symmetry of the $\alpha$-Pu phase.

Based on the strained-SP-GGA results for $\alpha$- and $\delta$-Pu,  the point defect properties of the two phases stand in stark contrast to one another. The vacancy in $\delta$-Pu has a relatively high formation enthalpy and a large negative relaxation volume, but the vacancy in $\alpha$-Pu is energetically much more favored with a small relaxation volume there. The opposite is the case for the self-interstitial defects, i.e. formation energies and volumes are smaller in the $\delta$-Pu phase than in $\alpha$-Pu. 

Point defects in both phases have quite unexpected properties and cannot easily be placed in categories with known materials of similar crystal structures. In the case of $\delta$-Pu, comparison with close-packed transition metals such as e.g. Cu, reveals sharp contrast. Cu has a small negative vacancy relaxation volume $\Delta V_R^{\text{Cu}} \approx -0.3$ at. vol., and a relatively large and positive self-interstitial relaxation volume $\Delta V_R^{\text{Cu}} \approx 1.9$ at. vol. Quite surprisingly, these properties resemble those of the point defects in the far-from close-packed $\alpha$-Pu phase. Consequently, the $\alpha$-Pu phase is expected to have a relatively high swelling bias within the conventional theory, comparable to that of Cu or Al, while $\delta$-Pu is predicted to have a greatly reduced swelling bias. We will discuss the prospect of radiation-induced aging in more detail in Sect.~\ref{sect:discussion}.
 
These properties can be rationalized within the spin/orbital DFT picture by the degree the $f$-electron manifold is influenced by bonding or anti-bonding states. We use the concept of formation spin moments Eq.~\ref{eq:fmom}, introduced in Sect.~\ref{sect:thermo}, to quantify the role of $f$-electron character in determining the unusual point defect properties found in $\alpha$- and $\delta$-Pu. It will be shown in the following sections that point defects in $\delta$-Pu reduce the magnitude of spin/orbital polarization in their vicinity. Hence in contrast to regular close-packed metals, the vacancy and the interstitial in this Pu phase can be accommodated with very small formation volumes. In contrast, point defects in $\alpha$-Pu are found to strongly increase the magnitudes of spin/orbital moments in their neighborhoods, which leads to large formation volumes.

\subsubsection{Modified SP-GGA for improved predictions of defect properties}
\label{sec:mod-sp}
Before embarking on a detailed discussion of point defect properties in $\alpha$- and $\delta$-Pu, we make a brief digression in this section on how collinear SP-GGA calculations can be improved upon to better reproduce the SP+SO+OP-GGA  results. This is important because accurate predictions of point defect energies and structures require relatively large supercells containing more than 100 atoms, and non-collinear SP+SO+OP-GGA calculations can become computationally challenging to converge to the levels needed for adequate determination of relaxation volumes. In contrast, collinear SP-GGA calculations are computationally expedient and straightforward to converge to high accuracy using state-of-the art computer resources and algorithms. However, SP-GGA underestimates the equilibrium volume by nearly $8\%$ for the $\alpha$ as well as the $\delta$-Pu phases.  While this is not an unreasonably large error, one should bear in mind that bonding in Pu metal is quite sensitive to atomic density, as it undergoes six structural phase transformations involving large volume changes within a temperature range of no more than 900~K. More importantly, as discussed in Sect.~\ref{sec:puredeltaPu}, collinear spin-polarized calculations predict a layered antiferromagnetic order with tetragonal symmetry to be the lowest-energy spin configuration for $\delta$-Pu. This breaks the cubic symmetry of the fcc phase. Coincidentally, at the SP+SO+OP-GGA equilibrium volume, which is within 1\% of experiment, the degree of tetragonality is quite small, less than $1\%$. It increases with increasing density, so that at the SP-GGA zero-pressure volume, it becomes as large as $\approx 6\%$, whereupon non-negligible errors are observed in predictions of defect structures and energies, as will be discussed in Sect.~\ref{sec:vacancydelta}.

Hence one may expect that the predictions made by SP-GGA for the formation energies/volumes of point defects in e.g. $\delta$-Pu can be brought to reasonable agreement with SP+SO+OP-GGA if they are performed at the SP+SO+OP-GGA equilibrium density. In practice, it turns out that the best way to conduct these modified SP-GGA calculations is to perform them at negative hydrostatic pressure $P$. This amounts to augmenting the SP-GGA exchange-correlation functional with a $PV$ term, where $V$ is the supercell volume that is allowed to relax variationally. This term can be thought of as mimicking the increased anti-bonding character of the occupied states due to additional symmetry breaking by spin-orbit coupling and orbital polarization. Of course, it is only a homogeneous term, and cannot account for local interactions that explicitly originate from orbital ordering. These interactions lead to the minor differences found in the predictions of the structural energetics of $\delta$-Pu predicted by the (001) and (110) spin/orbital polarizations, see Tab.~\ref{r-tab:1}. The external pressure $P$ is expected to depend (weakly) on the overall structure. It is found to be $P=-3$~GPa in the $\delta$-Pu phase, and $P=-5$~GPa in $\alpha$-Pu. These pressures are obtained within SP-GGA for the $\alpha$- and $\delta$-Pu structures at their respective SP+SO+OP-GGA equilibrium densities, with the spin and orbital moment vectors pointing in the (001) direction.

To our knowledge, the modified SP-GGA method discussed above has not been generally used to handle the relatively large tetragonal distortion predicted by SP-GGA at the theoretical equilibrium. The most commonly adopted way to handle this problem has been through imposition of a cubic shape constraint on the defect supercells \cite{Hernandez_2014,deltaDefects:HERNANDEZ:2017}. It is shown in the Appendix that in the dilute limit, such defect calculations yield identical results to supercell calculations, in which a non-hydrostatic (tetragonal) external stress is imposed. The magnitude of the stress $\sigma$ is chosen such that the perfect bulk fcc-Pu lattice becomes a stable equilibrium configuration. This amounts to augmenting the SP-GGA exchange-correlation functional with a $\sigma \eta$ term, where $\eta$ is the degree of tetragonality (e.g. $c/a$-ratio of an fct lattice), and is allowed to relax variationally. The generality of such a correction to the exchange-correlation functional is questionable, since it is directly dependent on the particular antiferromagnetic spin order chosen for the calculations. It does not have any direct relation with the SO and OP contributions that are missing in the SP-GGA calculations.

Finally, it should be noted that structural relaxations induced by point defects in lattices can contain substantial long-range components, which can lead to spurious interactions between periodic images. They can lead to slow convergence of calculated defect properties with supercell size. In Appendix, we show that finite supercell-size errors can be reduced when the supercell calculations are conducted in an external stress field, i.e. the free-energy functional is augmented with a $PV$ or a $\sigma \eta$ term. This gain in accuracy is relative to defect calculations in which supercell volumes or shapes are kept fix.

\subsubsection{Vacancies in $\delta$-Pu}
\label{sec:vacancydelta}

Vacancy formation enthalpies and compressibilities as well as relaxation volumes in $\delta$-Pu are given for SP-GGA and SP+SO+OP-GGA approximations in Tab.~\ref{r-tab:deltavac}.  All calculations were performed in 108-atom supercells with a $3\times3\times3$ K-point grid. All SP-GGA calculations are considered converged when atomic forces are reduced to $ < 0.002$~eV/\AA{} and elements of the stress tensor are converged to within 0.01-0.02 kBar. The corresponding convergence criteria for SP+SO+OP-GGA is typically 0.004~eV/\AA{} and 0.04 kBar. Structural parameters (including cell shapes) are always relaxed without constraint using the standard VASP procedure. For comparison, point defect properties in fcc Cu were also calculated in 108-site supercells and are listed in Tab.~\ref{r-tab:deltavac}. The VASP-PAW scheme and the PBE exchange-correlation functional with a plane-wave energy cutoff of 600~eV and a K-point grid of $4x4x4$ have been used for these calculations.

As was mentioned in Sect.~\ref{sect:methodology}, a variational formulation of SP+SO+OP-GGA is necessary for accurate calculations of atomic forces, and requires non-collinear magnetism derived from spinor wave functions. To our knowledge, fully variational SP+SO+OP-GGA calculations have in the past been only applied to perfect crystalline phases of Pu. No issues have arisen, because these systems stay collinear when initialized in a collinear state. This is not necessarily true for point defects, which may induce explicit non-collinearities in their local neighborhoods. In the present work, we choose to limit the scope to collinear spin configurations, while the calculations are fully variational using spinor wave functions. In order to maintain collinearity a quadratic penalty functional is added to the total-energy functional, which constrains the spin orientations along prescribed directions. In this way, vacancy energies, structures and compressibilities are studied in $\delta$-Pu with collinear AF-L10 spin order, polarized in two different spatial directions: (001) and (110). Certainly, much remains to be studied regarding coupling of non-collinear magnetic order to structural relaxations, which we defer to future works.

\begin{table}{}
\caption{ Formation energies, relaxation volumes, and the two contributions to formation compressibility (see Eqs. \ref{eq:beta}-\ref{eq:betaL}) of point defects,  in 108-site supercells of $\delta$-Pu, calculated within different levels of theory.
In each of the examples, the magnetic order is initialized in the usual L10 layered AF configuration and allowed to relax during the approach to electronic self-consistency as well as ionic relaxation. The final spin order in each case is found to be largely unchanged, in particular beyond and in the immediate vicinity of the point defects, and thus the AF-L10 can be considered metastable. Note that the monoclinic vacancy is not stable in dilated (-3 GPa) SP-GGA calculations. For comparison, the table also contains defect formation energies, relaxation volumes and compressibilities of the vacancy and the (001)-dumbbell self-interestitial in fcc copper at 0~GPa, calculated within PBE-GGA in 108-site supercells.   
}
\centering
\begin{tabular}{c c c c c c c c}
\hline\hline
SP-GGA              &Formation                       & Relaxation vol                                       &  $\Delta \beta_0$             &    $\Delta \beta_L$   \\ 
 (0 GPa)  &       energy (eV)                   & /atomic vol.                                           & (GPa)$^{-1}$            &     (GPa)$^{-1}$              \\[1ex]
\hline
\multicolumn{1}{l} {vac mono}                                         & 1.24                                & -2.33  \comnt{my data}           &  0.14    &  -0.052                            \\ 
\multicolumn{1}{l} {vac tetr}                                         & 1.42                                & -1.32  \comnt{my data}           &  0.022    &   -0.03                         \\ 
\multicolumn{1}{l} {int oct}                                     & 0.6                             & 1.00  \comnt{my data}           &  0.4                      &        0.022  \\ [1ex]
\hline\hline
SP-GGA    &&&&\\
 (-3 GPa) &&&&\\[1ex] 
\hline
\multicolumn{1}{l} {vac tetr}                                         & 1.47                                & -0.85  \comnt{my data}           &  0.25             &        -0.024           \\ 
\multicolumn{1}{l} {int (111)}                                     &   1.35                              &  1.32 \comnt{my data}           &    -                      &     -   \\ 
\multicolumn{1}{l} {int oct}                                     & 0.50                                & 1.32  \comnt{my data}           &  0.11              &  0.037                 \\ [1ex]
\hline\hline
SP+SO+OP+GGA        &&&& \\ 
(001)-polarization  &&&& \\[1ex] 
\hline
\multicolumn{1}{l} {vac tetr}                                         & 1.5                                & -0.94  \comnt{my data}           &  0.11        &  -0.022                        \\ 
\multicolumn{1}{l} {int oct}                                     & 0.67                                & 0.85  \comnt{my data}           &  0.046      &   0.02                          \\ 
\hline\hline
SP+SO+OP-GGA       &&&&   \\  
(110)-polarization  &&&& \\[1ex] 
\hline 
\multicolumn{1}{l} {vac tetr}                                      & 1.56                                & -1.0  \comnt{my data}           &  0.19         & -0.026                       \\ \multicolumn{1}{l} {int oct}                                   & 0.98                                & 1.2  \comnt{my data}           &  -0.034             &   0.03 \\
\hline\hline
fcc-Cu        &&&& \\ 
GGA (0 GPa)  &&&& \\[1ex] 
\hline
\multicolumn{1}{l} {vac}                                         & 1.1                                & -0.33  \comnt{my data}           &  -0.006        &  -0.0024                        \\ 
\multicolumn{1}{l} {int (001)-dumbbell}                                     & 3.0                                & 1.87  \comnt{my data}           &  0.007      &   0.0014                          \\ 
\hline
\end{tabular}
\label{r-tab:deltavac}
\end{table}

\begin{figure}[t]
\includegraphics[width=0.95\columnwidth]{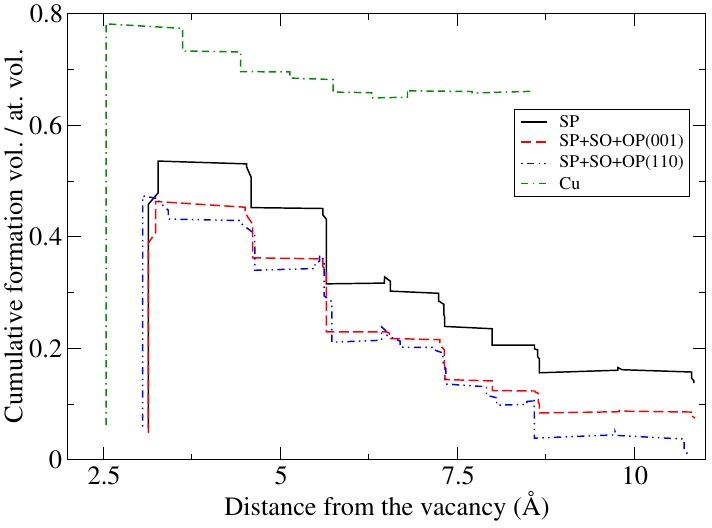}
\caption{Cumulative formation volume $\Delta V_F^C(d)$ as a function of distance $d$ from a vacant site in Cu (green dashed-dotted line) and in $\delta$-Pu calculated within SP-GGA (black solid line), SP+SO+OP-GGA polarized along (001) (dashed red line), and SP+SO+OP-GGA polarized along (110) (blue dashed-dotted line).  The plots depict the accumulated volume change as a function of distance from the vacancy in the defect supercells. The cumulative formation volumes are given in units of per-atom equilibrium volume of the respective perfect crystal, }
\label{fig:vorodist}
\end{figure}

\begin{table}{}
\caption{ Formation spin moments of intrinsic point defects in 108-site supercells of $\delta$-Pu, initialized in the L10-AF spin configuration, and calculated within different levels of theory. The formation moments are listed in fractions of spin-moment-per-atom in the perfect lattice.  
}
\centering
\begin{tabular}{c c c c c c c}
\hline\hline
&  &SP                       & & SP+SO+OP     &   &   SP+SO+OP                                                            \\
&  & -3 GPa                  &  & (001)-polarization & & (110)-polarization  \\[1ex]
\hline
\multicolumn{1}{l} {vac tetr}  &        &       -0.17   &                            &             -0.22         &   &  -0.36   \\ 
\multicolumn{1}{l} {int oct}     &                                &   -0.65 &                               &  -1.15  &    & -0.83   \\
\hline
\end{tabular}
\label{r-tab:deltamom}
\end{table}

\begin{figure}[t]
\includegraphics[width=0.95\columnwidth]{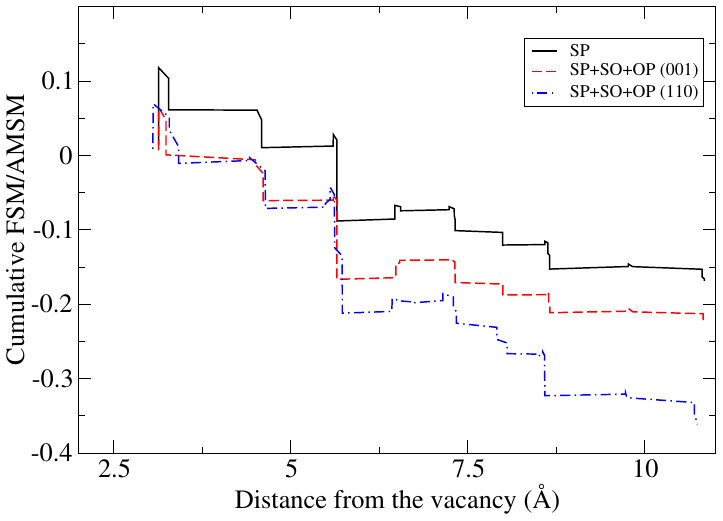}
\caption{Cumulative formation spin moment $\Delta \Sigma_F^C(d)$ as a function of distance $d$ from a vacant site in $\delta$-Pu. It is the accumulated induced spin moment within a distance $d$ from the vacant site. It is given in the units of the average magnitude of per-atom spin moments (AMSM) in $\delta$-Pu, calculated within SP-GGA in black, SP+SO+OP-GGA polarized along (001) in dashed red, and SP+SO+OP-GGA polarized along (110) in dashed-dotted blue.}
\label{fig:2}
\end{figure}

\begin{figure}
\includegraphics[width=\linewidth, trim= 1 75 100 20]{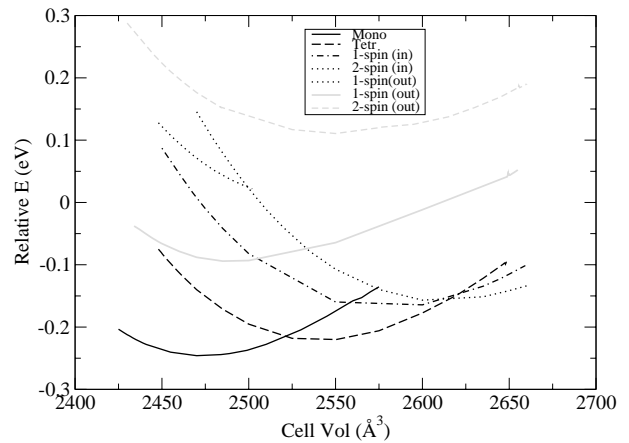}
\caption{Vacancy formation energies in $\delta$-Pu  versus volume for different spin configurations, as described in the text and Table~\ref{r-tab:flips}.  A quadratic least squares fit to all configurations versus volume has been subtracted to highlight the differences between the curves. The curves all terminate  at their lower ends near zero pressure (theoretical equilibrium) and at -30 kbar at the upper end (the stress conditions that match the experimental bulk volume), the two limits where DFT results are commonly reported for Pu.  The character of bonding, structural relaxation, and magnetic order of the defect changes significantly over the different densities.}
\label{r-fig:flips}
\end{figure}

In the past, most reported calculations of the vacancy in $\delta$-Pu were performed using collinear SP-GGA theory, with the spin density initialized in a layered L10 AF spin order, and the atomic density constrained to the theoretical zero-pressure equilibrium. Under these conditions, SP-GGA predicts large static moments on each Pu site, which in turn induces the bulk $\delta$-Pu to have substantial tetragonal distortion, see Tab.~\ref{r-tab:1}. In order to correct for this error, cubic shape constraints were then imposed on the defect supercells.  The most elaborate calculations have been conducted by Hernandez et al.\cite{spinDelta:HERNANDEZ:2019}, who found the lowest-energy vacancy to have a distorted monoclinic structure, which we denote by ``vac mono'' in Tab.~\ref{r-tab:deltavac}. The calculations that we report in this paper however, differ from that previous work \cite{spinDelta:HERNANDEZ:2019} in that they do include full relaxations of supercell shapes. This reduces finite supercell-size errors on the calculated properties, but does not include any correction for the erroneous tetragonal distortion predicted by SP-GGA.  Nevertheless, the ``vac mono'' remains the ground-state vacancy structure at the theoretical equilibrium density even when full supercell shape relaxations are employed. Table~\ref{r-tab:deltavac} records the resulting thermodynamic properties of ``vac mono'' defect. We find a very large relaxation volume, corresponding to a lattice contraction of more than two atomic volumes per vacant site. However, this vacancy configuration becomes energetically less favorable relative to other vacancy species as the lattice is expanded, see Fig.~\ref{r-fig:flips}.  It is dynamically unstable near the experimental ambient-pressure density. Therefore, it cannot exist at these densities. Instead, a different vacancy configuration denoted by ``vac tetr'' in Tab.~\ref{r-tab:deltavac}, becomes the ground-state vacancy structure in $\delta$-Pu, see Fig.~\ref{r-fig:flips}. It can be obtained by simply removing an atom from a 108-atom supercell of perfect bulk $\delta$-Pu, with the spin density initialized in the L10-AF spin configuration, whereupon electronic degrees of freedom are brought to self-consistency, followed by relaxations of the ionic degrees of freedom as well as the supercell shape. The resulting formation energies and relaxation volumes within both SP- and SP+SO+OP-GGA are listed in Tab.~\ref{r-tab:deltavac}. As has been argued earlier in this paper, we expect the most reliable collinear theory for point defect properties in $\delta$-Pu to be the (001)-polarized SP+SO+OP-GGA, which comes at a relatively high computational cost. Nevertheless, the overall agreement with the simpler SP-GGA strained to $-3$~GPa is reasonable. Furthermore, for bulk $\delta$-Pu, both theories predict rather small tetragonal distortions, less than $1\%$, in contrast to SP-GGA at theoretical zero-pressure density, which predicts 6 times larger distortion, see Tab.~\ref{r-tab:1}.     

The relaxation volumes recorded in Tab.~\ref{r-tab:deltavac} for ``vac tetr'' reveal that even this vacancy induces a large contraction in the $\delta$-Pu lattice on the order of one bulk-Pu atomic volume. This is several times larger than vacancy relaxation volumes in typical close-packed transition metals, such as fcc-Cu, where each vacancy contracts the lattice by no more than $1/3$ of an atomic volume. Likewise, the vacancy formation enthalpy is substantially larger $\approx 1.5$ eV in $\delta$-Pu as compared to $1.1$ eV in fcc-Cu, calculated within GGA. This implies a much smaller equilibrium vacancy concentration in $\delta$-Pu than in a typical fcc metal.  

It should be noted that in spite of the relatively small tetragonality error of the dilated SP-GGA as well as SP+SO+OP-GGA theories for bulk $\delta$-Pu, the predicted structural relaxations about the vacancy contain non-negligible anisotropy. This is a result of the symmetry of the underlying static L10 AF spin order, which consists of ferromagnetic (001) layers, with adjacent planes having opposite spin directions. As a result, in the fcc lattice, each Pu atom is surrounded by 8 nearest neighbors (NN) with antiparallel spin orientations, and four NN-sites with parallel spins. When a Pu atom is removed from a lattice site, both (001)-polarized SP+SP+OP-GGA and SP-GGA predict the eight NNs with antiparallel spins to relax inward by about 4\%, while (110)-polarized SP+SO+OP-GGA predicts a larger inward relaxation of 6.2\%. On the other hand, the four parallel-spin NNs relax inward within (001)-polarized SP+SO+OP-GGA by 1.8\%,  do not relax measurably within SP-GGA, and relax outward by more than 1.5\% according to (110)-polarized SP+SO+OP-GGA. This should be compared with the vacancy relaxation pattern in fcc Cu, where all 12 NNs move inward by about 1.4\%. It is evident that the (001)-polarized SP+SO+OP-GGA exhibits the least relaxation anisotropy, followed by SP-GGA. This is quite expected when comparing the predicted anisotropies by the different theories for the perfect $\delta$-Pu lattice, see Tab.~\ref{r-tab:1}.  In particular, it can be seen that the (110)-polarized SP+SO+OP-GGA exhibits twice the anisotropy for the perfect $\delta$-Pu lattice compared to the other two approximations, resulting in similarly larger anisotropy of the local relaxations around the vacancy.

Further examination of local relaxations around the vacancy reveals a fundamental and novel feature of interactions in plutonium metal. In crystalline solids, lattice defects cause structural relaxations that within the so-called core regions around the defects depend on the detail of atomic positions and interatomic interactions, while outside can be described by continuum elasticity in order to account for the response of the host lattice to the deformation from the core regions. The former are often expected to be confined to a few neighbor shells around the defects, while generating long-range elastic strain fields in the solid. The assumption that the non-linear core region can be contained to within several tens of atoms underlies the supercell technique for calculation of energies and structures of lattice defects from first principles.

In order to understand the nature of volume relaxations and the extent of the non-linear core region around the vacancy in $\delta$-Pu, we examine the spatial distribution of its dilatational strain field. This can be done by Voronoi decomposition of the defect supercells. The Cumulative formation volume $\Delta V_F^C(d)$ as a function of distance $d$ from the vacancy in a defect supercell containing $(N-1)$ atoms can be defined as
\begin{equation}
  \label{eq:dVc}
  \Delta V_F^C(d) = \sum_{i=1}^{N-1} \left(\omega_i^d-\Omega_L\right)~H(d-d_i),
\end{equation}
where $\omega_i^d$ is the Voronoi volume of atom $i$ in the defect supercell, $\Omega_L$ is the atomic volume in the perfect lattice, and $H(x)$ is the Heavyside function, with $H(x) = 0$ for $x<0$, and otherwise $H(x) = 1$. Hence $\Delta V_F^C(d)$ represents the contribution to the formation volume from a spherical region of radius $d$ around the vacancy. At large distances, the cumulative formation volume $\Delta V_F^C$ approaches the thermodynamic formation volume, $\Delta V_F$, defined in Eq.~\ref{eq:relV}. The radial extent $d_c$ of the non-linear core region can thus be defined as the distance, beyond which $\left|\Delta V_F^C(d_c)-\Delta V_F\right| < \epsilon$, with $\epsilon << 1$. This is because in an isotropic continuum elastic medium, the dilatational strain field due to a misfit inclusion can be shown to be a harmonic function, and thus the incremental volume change integrated over a spherical shell far from the inclusion is independent of its radius. As a result, reasonably well-converged estimations of formation volumes require only supercell sizes in which the nearest point-defect images are only about $2d_c$ apart. 

Assuming that the core regions are contained within a few neighbor shells of the point defects, supercells containing on the order of 100 atoms should be adequate for calculation of their properties. Figure~\ref{fig:vorodist} shows the distribution of $\Delta V_F^C(d)$ as a function of distance $d$ from the vacancy, for the three levels of theory studied here. The reader should be reminded here that Fig.~\ref{fig:vorodist} depicts formation volumes $\Delta V_F$, while Tab.~\ref{r-tab:deltavac} lists relaxations volumes $\Delta V_R$. For the vacancy, $\Delta V_R = \Delta V_F - 1$, see Eqs.~\ref{eq:relV} and ~\ref{eq:relV1}. It is thus apparent from Fig.~\ref{fig:vorodist} that if the volume relaxations were confined to the first-neighbor shell, as they usually are around the vacancy in standard metallic systems, the relaxation volume in $\delta$-Pu would be no less than -0.5. However, volume relaxations beyond the first neighbor shell nearly double the magnitude of $\Delta V_R$.  The three theories applied to $\delta$-Pu vacancy agree on the extent of the non-linear core region being larger than can be comfortably contained within 107-atom supercells used in this study.  For comparison, we also show in Fig.~\ref{fig:vorodist}. the distribution of $\Delta V_F^C(d)$ for the vacancy in fcc Cu. It is apparent that in this system, the non-linear core region can be considered confined to within a few neighbor shells of the vacancy, and the 107-atom supercell is thus expected to make an adequate representation of the volume relaxations in this system. On the other hand, for the $\delta$-Pu vacancy, it is reasonable to expect based on the nearly monotonous drop in $\Delta V_F^C$ at large distances in Fig.~\ref{fig:vorodist}, that vacancy calculations in larger supercells will yield further increase in the magnitudes of the calculated relaxation volumes.

The origin of the unusual properties of the vacancy in $\delta$-Pu can be traced to the bonding characteristics of the $f$-electrons in this phase. In the simplest band picture, such as non-magnetic DFT, there is a high density of narrow $f$-electron bands at the Fermi level for high-symmetry phases such as $\delta$-Pu. Broken-symmetry phases (e.g., structural ones like $\alpha$-Pu) can lift this degeneracy and lower the overall energy of the system.  When instead allowing for symmetry-breaking spin/orbital polarization, sizeable spin/orbital moments form in e.g. $\delta$-Pu, that are largely concentrated within the atomic spheres due to increased occupation of anti-bonding states. They split the degenerate bands at the Fermi level and thereby lower the total-energy of the high-symmetry phases and stabilize them at expanded volumes. 

In order to conduct a more quantitative study of this mechanism, we introduced in Sect.~\ref{sect:thermo} a measure of $f$-electron bonding in Pu in terms of average magnitude of spin moments per atom (AMSM), see Eq.~\ref{eq:avmom}. For example, the AMSM in $\delta$-Pu at zero pressure, calculated within SP+SO+OP-GGA is $4.4~\mu_B$. Furthermore, for point defects in lattices, the formation spin moment (FSM) defined in Eq.~\ref{eq:fmom} can provide a quantitative measure of the effect of lattice defects on $f$-electron bonding in their neighborhoods. The FSM for vacancies in $\delta$-Pu in units of AMSM of $\delta$-Pu are listed in Tab.~\ref{r-tab:deltamom}. We see that introduction of a vacant site in the $\delta$-Pu lattice reduces the spin moment magnitudes  of the surrounding $f$-electrons. While this effect is not very large, it can be shown that in combination with the weak bonding in $\delta$-Pu, it is the leading cause of the large and negative relaxation volume of the vacancy in $\delta$-Pu, see Tab.~\ref{r-tab:deltavac}. It is also responsible for the unusual structural relaxations around the vacancy in $\delta$-Pu with a wide non-linear core region, discussed above in conjunction with Fig.~\ref{fig:vorodist}.

In order to study the coupling of vacancy-induced spin polarization to the structural relaxations around the defect, we define in analogy with $\Delta V_F^C(d)$ above. see Eq.~\ref{eq:dVc}, the cumulative FSM $\Delta \Sigma_F^C(d)$ as a function of distance $d$ from the vacancy
\begin{equation}
  \label{eq:dFSMc}
  \Delta \Sigma_F^C(d) = \sum_{i=1}^{N-1} \left(\sigma_i^d-\Sigma_L\right)~H(d-d_i).
\end{equation}
Above $\sigma_i^d$ is the local spin moment magnitude of the $i$th atom in the defect supercell, and $\Sigma_L$ is the AMSM of the perfect lattice, and $H(x)$ is the Heavyside function. Note that at large distances,  $\Delta \Sigma_F^C$ approaches the thermodynamic FSM value. Figure~\ref{fig:2} shows the spatial distribution of $\Delta \Sigma_F^C(d)$ for the three levels of theory in this study. It can be seen that the Pu atoms in the first neighbor shell of the vacancy slightly increase their spin polarization, leading to a positive $\Delta \Sigma_F^C$ at small distances. This is reasonable considering that $f$-electron spin/orbital polarization is already saturated in $\delta$-Pu. However, the relatively weak bonding in $\delta$-Pu allows the atoms in the first neighbor shell around the vacancy to move substantially closer, whereupon the increased atomic disorder reduces the overall spin/orbital polarization leading to increased bonding and further contraction. As a result, beyond the first neighbor shell, $\Delta \Sigma_C^F(d)$ turns negative and drops nearly monotonously away from the vacancy, see Fig.~\ref{fig:2}. The similarity of the $\Delta \Sigma_F^C(d)$ and $\Delta V_F^C(d)$ distributions is apparent by inspection of Figs.~\ref{fig:vorodist} and~\ref{fig:2}. Hence, the combination of weak bonding in $\delta$-Pu and negative FSM values causes an anomalously large negative relaxation volume $\Delta V_R$ for the vacancy in $\delta$-Pu. In other words, the formation volume $\Delta V_F$ of the vacancy in $\delta$-Pu (note: $\Delta V_F = \Delta V_R + 1$) is nearly zero, which means injection of vacancies into a $\delta$-Pu metal bar leads to little measurable change of its dimensions. 

Table~\ref{r-tab:deltavac} also records the formation compressibilities of vacancies calculated within different levels of theory, as defined in Sect.~\ref{sec:compress}. The total compressibility $\Delta \beta$ of the vacancy is composed of two contributions: (i) $\Delta \beta_0$, see Eq.~\ref{eq:beta0}, which except for the outlier ``vac tetr'' calculated within SP-GGA at theoretical equilibrium, attains quite large values ranging from 0.11 to 0.25~(GPa)$^{-1}$, and measures the explicit effect of lattice softening by the vacancy, and (ii) $\Delta \beta_L$, see Eq.~\ref{eq:betaL}, which is about 5 to 10 times smaller than $\Delta \beta_0$ and measures the vacancy compressibility in the absence of any defect-induced change in the crystal's elasticity. It should be noted that these calculations are numerically quite difficult to converge, in particular in the presence of spin/orbital polarization. Nevertheless, we find the different theories to be consistent with strong lattice softening caused by the vacancies in $\delta$-Pu leading to relatively large compressibility of this defect.

It is instructive to study the defect formation compressibilities in a typical transition metal in order to provide context for the calculations presented above for $\delta$-Pu. For this purpose, we have conducted detailed calculations of $\beta_0$ and $\beta_L$ for the point defects in fcc Cu, listed in Tab.~\ref{r-tab:deltavac}. It can be seen that the vacancy formation compressibilities in Cu are more than an order of magnitude smaller than in $\delta$-Pu. This is partly due to the much stiffer lattice as reflected in the calculated bulk modulus of 136.4 GPa for Cu, which is more than three times larger than the value for $\delta$-Pu. Nevertheless, the largest contribution does clearly originate from vacancy-induced changes to the local $f$-electron correlations and bonding causing softening of the $\delta$-Pu lattice bulk modulus.

\begin{table}{}
\caption{ Effect of low-energy spin flips in the vicinity of the vacancy in $\delta$-Pu on its properties. Calculations are performed in 108-site supercells of $\delta$-Pu, within SP-GGA.
In each of the examples, the magnetic order is initialized in the L10-AF configuration with additional spin flips imposed on the Pu atoms neighboring the vacancy. Each case is found to be metastable during the iteration to self-consistency.
The symmetry-related magnetic degeneracy is given for each case along with defect volume and energy. Details of the initial spin orders are given in the text.
The reference, undefected bulk state is the usual L10 layered antiferromagnet.
The associated SP-GGA bulk modulus in the perfect crystal is 44.7 GPa at zero pressure and 36.0 GPa at -3 GPa.
}
\centering
\begin{tabular}{c c c c c c c c}
\hline\hline
0 GPa                                                           & Degen.           &Formation                       & Relaxation vol                                                                               \\ 
SP-GGA                                                        &                       & energy (eV)                   & /atomic vol.                                                                    \\[1ex]
\hline 
\multicolumn{1}{l} {vac one in }    \comnt{5p}           &           4            & 1.57  \comnt{OK}               & - 0.95  \comnt{OK}                                               \\
\multicolumn{1}{l} {vac one out}                               &           8            & 1.42 \comnt{OK}            & - 2.00  \comnt{OK}                                                \\
\multicolumn{1}{l} {vac two in}     \comnt{6p}          &           2            & 1.70                                  & -0.44                                                                             \\ 
\multicolumn{1}{l} {vac two out}   \comnt{4}            &           4            & 1.72  \comnt{OK}              & - 1.56    \comnt{OK}                                           \\
\hline\hline
-3 GPa                                                           & Degen.           &Formation                       & Relaxation vol                                                                  \\ 
SP-GGA                                                         &                       & energy (eV)                   & /atomic vol.                                                                \\[1ex]
\hline
\multicolumn{1}{l} {vac one in }                                &           4            & 1.44  \comnt{OK}           & - 0.57  \comnt{OK}                                        \\
\multicolumn{1}{l} {vac one out}                               &           8            & 1.60 \comnt{OK}            & - 0.73  \comnt{OK}                                \\
\multicolumn{1}{l} {vac two in}   \comnt{6p}             &           2            & 1.41 \comnt{OK}            & - 0.19                                             \\ 
\multicolumn{1}{l} {vac two out}                               &           4            & 1.74  \comnt{OK}           & - 0.52   \comnt{OK}                                \\ [1ex]
\hline
\end{tabular}
\label{r-tab:flips}
\end{table}

We conclude this section by a discussion of alternate spin arrangements and their effect on the properties of the vacancy in $\delta$-Pu. To our knowledge, all studies of point defects in $\delta$-Pu in the past have assumed that the spin configuration of the defect lattice does resemble the perfect lattice and therefore is initialized with the usual L10 layered AF order. Above, we have denoted this particular vacancy state by ``vac tetr''. In addition to this default spin order, we also have examined SP-GGA calculations initialized with four other spin configurations, where the spin order in the neighborhood of the vacant site is altered. They are: (i) the "one in" case, which reverses a single, in-plane nearest neighbor spin moment in the same (001) plane as the vacancy; (ii) the "two in" case, which reverses a pair of in-plane moments lying on opposite sides of the vacancy; (iii)
the "one out" case, which reverses a single, out-of-plane nearest neighbor moment in an adjacent (001) plane to the vacancy; (iv) the "two out" case reverses the moments of two neighbor spins on different (001) planes on opposite sides of the vacancy. 
In practice, subsequent iterations to full self-consistency preserve these magnetic structures, indicating metastability.
The resulting vacancy energies are given in Tab.~\ref{r-tab:flips} and their relative order displayed in Fig.~\ref{r-fig:flips} at a range of volumes. It can be seen that at -3~GPa tension, SP-GGA predicts both ``vac one in'' and ``vac two in'' vacancies to be slightly lower in energy than ``vac tetr''. The magnitudes of their relaxation volumes, in particular that of ``vac two in'', are significantly smaller than ``vac tetr'', i.e. they lead to much less contraction of the surrounding lattice. This difference can be understood by noting that the spin flips that constitute the excited vacancy configurations ``vac one in'' and ``vac two in'' increase the local ferromagnetic order around the vacant site, i.e. the number of nearest neighbors with parallel instead of antiparallel spins. The ferromagnetic order expands the lattice relative to the AF order, but it is energetically less favorable in bulk $\delta$-Pu. 

We thus conclude that low-energy variations in the local magnetic order around the vacancy within SP-GGA can reduce the defect relaxation volumes substantially.
However, while spin flip excitations such as in ``vac one in'' and ``vac two in'' actually lower the vacancy energy in SP-GGA relative to the default L10 configuration ``vac tetr'', they are energetically unfavorable in SP+SO+OP-GGA.
This is observed in comparative SP-GGA and SP+SO+OP-GGA calculations in smaller 31-atom supercells containing a vacancy,  where the spin reversals that lower the total energy within SP-GGA, raise the total energy according to SP+SO+OP-GGA. 
This discrepancy can be attributed to increased magnetic exchange coupling within the SP+SO+OP-GGA, which raises the energy of spin-parallel nearest neighbors relative to spin-anti-parallel ones. Thus the evidence at this time is that the lowest energy vacancy configuration within SP+SO+OP-GGA retains the unaltered L10-AF structure from the perfect bulk crystal.

The above discussion suggests the importance of spin fluctuations in determining finite-temperature properties of  point defects in $\delta$-Pu, potentially leading to non-Arrhenius behavior. In fact, electronic excitations are known to cause the anomalous Invar-like effect observed in bulk $\delta$-Pu.\cite{LAWSON:INVAR:2002,Migliori:PNAS:2016} They may also alter the kinetics of void swelling by affecting the point defect interactions. This occurs when spin fluctuations adjust the relaxation volumes, as described above for spin-excited vacancies, and thereby change the associated strain energies. As a proof of principle, in Sect.~\ref{sect:discussion}, we model the possible effects of spin fluctuations on radiation aging behavior, albeit somewhat crudely, by computing the thermal equilibrium formation energy and volume of the vacancy in $\delta$-Pu in SP-GGA under -3 GPa pressure. For this purpose, we recognize that by symmetry, there are multiple equivalent sites for each of the spin flips described above, which results in degeneracies listed in Tab.~\ref{r-tab:flips}.

In summary, the vacancy in $\delta$-Pu differs markedly in its structure and energy from standard close-packed metals. Its equilibrium concentration is anomalously low, and it induces strong lattice contraction in its neighborhood. This is a result of the weak bonding in $\delta$-Pu combined with the vacancy having an overall negative influence on the  $f$-electron spin/orbital polarization in the $\delta$-Pu lattice. Consequently, large local contraction around the vacancy can occur at relatively low energy cost. In the final analysis, it is found that the dilated SP-GGA approximation can produce results in qualitative agreement with the SP+SO+OP-GGA calculations. However, when studying more subtle issues, such as the precise relative order of low-energy spin excitation, the two approximations can differ significantly. Nevertheless, it is shown that low-energy localized moment excitations can cause anomalous temperature-dependent properties of point defects in Pu. At this point, our knowledge of the nature of these fluctuations is wanting, and most research and development in this direction still remains to be done.

\bigskip
\subsubsection{Self-interstitials in $\delta$-Pu}

\begin{figure}
\includegraphics[width=0.95\columnwidth]{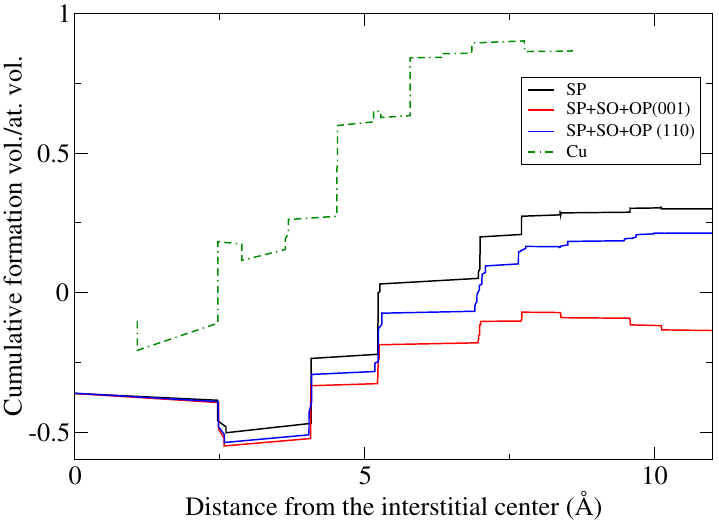}
\caption{Cumulative formation volume $\Delta V_F^C(d)$ as a function of distance $d$ from a self-interstitial center in Cu (green dashed-dotted line) and in $\delta$-Pu calculated within SP-GGA (black solid line), SP+SO+OP-GGA polarized along (001) (dashed red line), and SP+SO+OP-GGA polarized along (110) (blue dashed-dotted line).  The plots depict the accumulated volume change as a function of distance from the center of the self-interstitial in the defect supercells. The cumulative formation volumes are given in units of per-atom equilibrium volume of the respective perfect crystal, Note that the interstitial center in $\delta$-Pu is the octahedral site, while in fcc-Cu it is at the midpoint of the 001-dumbbell. }
\label{fig:3.5}
\end{figure}

\begin{figure}
\includegraphics[width=0.95\columnwidth]{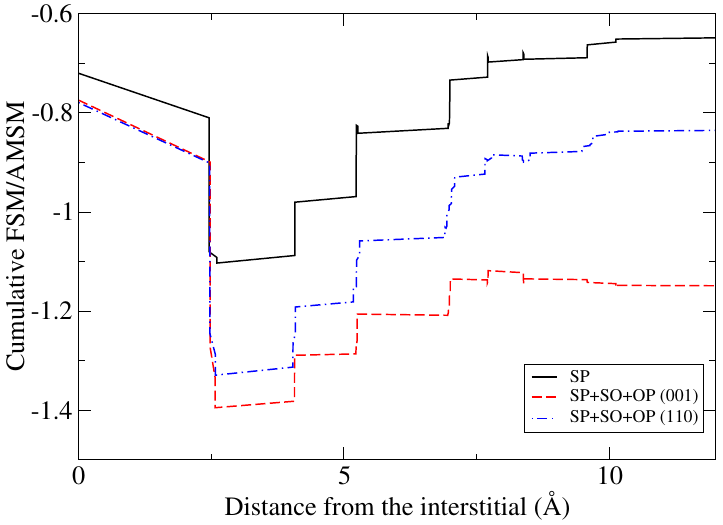}
\caption{Cumulative formation spin moment $\Delta \Sigma_F^C(d)$ as a function of distance $d$ from the octahedral self-interstitial in $\delta$-Pu. It is the accumulated induced spin moment within a distance $d$ from the octahedral interstitial site. It is given in units of the average magnitude of per-atom spin moments (AMSM) in $\delta$-Pu, calculated within SP-GGA in black, SP+SO+OP-GGA polarized along (001) in dashed red, and SP+SO+OP-GGA polarized along (110) in dashed-dotted blue.}
\label{fig:3}
\end{figure}

Typical fcc materials have distinct interstitial defect configurations commonly known as split dumbbell and the octahedral site defects. In the split dumbbell configuration, the interstitial atom forms a dimer with a lattice atom. The dimer bond is usually directed along either (111) or (001) directions. In $\delta$-Pu, the (111)-dumbbell is high energy, while the (001)-dumbbell is unstable within both SP-GGA and SP+SO+OP-GGA, and relaxes into the octahedral configuration. The interstitial energies in $\delta$-Pu, calculated using 109-atom supercells using a $3\times 3\times 3$ K-point grid are reported in Tab.~\ref{r-tab:deltavac}.\cite{deltaDefects:HERNANDEZ:2017,magDefects:HERNANDEZ2020} The octahedral interstitial is an extra atom in the fcc lattice residing half-way along a cube side between two next nearest-neighbor sites. It has six nearest neighbors. In the AF-L10 spin configuration, the cubic symmetry of the perfect $\delta$-Pu lattice is broken, and the six degenerate NN distances break up into two groups of four and two. Within dilated (-3 GPa) SP-GGA as well as 001-polarized SP+SO+OP-GGA, the empty octahedral site in the perfect $\delta$-Pu lattice is surrounded by four NNs at $d_1 = 2.30$~\AA , and two NNs at $d_2 = 2.32$~\AA ~distance, while within (110)-polarized SP+SO+OP-GGA $d_1 = 2.29$~\AA ~and $d_2 = 2.33$~\AA . Upon incorporation of an interstitial atom in the octahedral site, the nearest neighbor sites move outwards with $d_1(d_2) = 2.46(2.61)$~\AA ~ within dilated SP-GGA, $d_1(d_2) = 2.48(2.58)$~\AA ~ within (001)-polarized SP+SO+OP-GGA and $d_1(d_2) = 2.48(2.59)$~\AA {} within (110)-polarized SP+SO+OP-GGA. Hence the nearest-neighbor relaxations around the octahedral self-interstitial in $\delta$-Pu are predicted to be very similar within the different levels of theory, with moderate signature of anisotropy inherited from the L10-AF spin configuration.

In typical closed-packed metals, such as Cu, the self-interstitial is a very high-energy defect. We have calculated the properties of the (001)-dumbbell self-interstitial in Cu, using 109-site supercells and a $4\times 4\times 4$ K-point grid. The results are listed in Tab.~\ref{r-tab:deltavac}. The self-interstitial formation enthalpy is indeed very high $\sim 3.0$ eV, and the relaxation volume of $1.9$ atomic volumes is quite sizable.  The story is quite the opposite for $\delta$-Pu. Table~\ref{r-tab:deltavac} reports the formation enthalpies and the relaxation volumes of the octahedral self-interstitial in $\delta$-Pu, calculated within SP-GGA as well as SP+SO+OP-GGA. These approximations predict a surprisingly low formation energy, clearly less than $1$~eV, and a formation volume that is close to only 1 atomic volume. More precisely, the formation volume calculated within dilated SP-GGA is 1.32, and reduces to $0.94$ when SO+OP corrections are included in the (001)-polarization.

It is interesting to compare the cumulative formation volumes $\Delta V_F^C(d)$, see Eq.~\ref{eq:dVc}, for the octahedral interstitial in $\delta$-Pu and the (001)-dumbbell interstitial in fcc Cu. Figure~\ref{fig:3.5} shows the dependence of $\Delta V_F^C$ on distance away from the respective self-interstitial centers in Cu and in $\delta$-Pu calculated within three levels of theory. It should be noted that for the octahedral site in $\delta$-Pu, the interstitial center is the octahedral site itself, while for the dumbbell interstitial in fcc Cu, it is the midpoint of the dumbbell. Furthermore, the reader should be reminded that Tab.~\ref{r-tab:deltavac} reports relaxation volumes $\Delta V_R$, while the asymptotic value of $\Delta V_F^C(d >> 1)$ is the formation volume $\Delta V_F$. For interstitials $\Delta V_F = \Delta V_R - 1$, see Eqs.~\ref{eq:relV} and \ref{eq:relV1}. Inspection of the $\Delta V_F^C$ distributions in Fig.~\ref{fig:3.5} reveals that in Cu, the two interstitial atoms are compressed by about 10\% each. It can thus be argued that the dumbbell generates a deformation corresponding to addition of 0.8 fraction of a Cu atom in the lattice. In response, the neighboring shells expand since they cannot pack as efficiently as in the pristine fcc lattice. The non-linear tensile strain field emanating from the dumbbell is quite long-range and as a result, the non-linear structural relaxations induced by the self-interstitial in Cu are not fully contained within a 109-atom supercell, see Fig.~\ref{fig:3.5}. This can be deduced because, as explained in Sect.~\ref{sec:vacancydelta}, in the far-field linear-elastic relaxation regime, the integrals of the displacement field from a point defect over spherical shells are independent of the sphere radii. In other words, in the linear-elastic regime, $\Delta V_F^C(d)$ becomes independent of distance $d$. In contrast to Cu, not only the octahedral interstitial but also its six nearest neighbors become compressed, which sums up to a total of 50\% compression. Hence the octahedral interstitial and its six nearest neighbors generate a deformation corresponding to addition of half a Pu atom in the $\delta$-Pu lattice. It can thus be observed in Fig.~\ref{fig:3.5} that the structural relaxations induced by the self-interstitial in $\delta$-Pu are clearly more short-range and better contained within a 109-atom supercell than is the case for the (001)-dumbbell interstitial in Cu. Note that the three theories completely agree on the size of the octahedral interstitial atom. The deviation in the calculated interstitial relaxation volume within SP-GGA ($\Delta V_R = 1.32$) from e.g. (001)-polarized SP+SO+OP-GGA ($\Delta V_R = 0.94$) is manifestly due to lattice relaxations away from the octahedral site. Below, we will relate this discrepancy to the differential predictions of the different theories for the  distribution of $f$-electron correlations induced by the octahedral interstitial in $\delta$-Pu lattice, 

The high formation energy of self-interstitials in Cu implies extremely low equilibrium concentrations, and the large relaxation volume implies very strong coupling to the stress field from vacancies, dislocations and other defects.  This imposes a persistent driving force towards recombination or absorption at sinks. As a result, when modeling radiation bombardment of close-packed metals, it suffices to neglect the thermal equilibrium bulk self-interstitial concentration.  There are effectively no interstitials present in the absence of irradiation.  Similarly, dislocations in Cu will absorb any interstitials irreversibly and climb strongly. In contrast, the low self-interstitial formation enthalpies and volumes for $\delta$-Pu imply that the bulk equilibrium concentration of self-interstitials in $\delta$-Pu is by far larger than vacancies, and furthermore, the fairly weak strain fields induced by interstitial-induced relaxation volumes interact no more strongly than the vacancies with sinks like dislocations and grain boundaries. Here, as compared to standard fcc metals, the self-interstitials are not preferentially absorbed (versus vacancies) at dislocations, there is little to no net dislocation climb expected, and excess interstitials will remain to efficiently annihilate excess vacancies. Hence, excess vacancies and voids generated as a result of radiation damage recombine much more effectively with self-interstitials in $\delta$-Pu than typically occurs in close-packed metals. This implies that swelling rates in $\delta$-Pu are significantly lower and incubation times much longer than expected for standard metals.

The calculations for self-interstitial formation enthalpies and volumes in $\delta$-Pu discussed hitherto, see Tab.~\ref{r-tab:deltavac},  assume static SP- and SP+SO+OP-GGA approximations.  They may change when allowing for spin fluctuations. Nevertheless, the self-interstitial remains most likely a far more favorable lattice defect in $\delta$-Pu than the vacancy. This can be rationalized by noting that the $\delta$-Pu phase, while close-packed in structure and symmetry, is in fact low-density relative to $\alpha$-Pu. Consequently, considering self-interstitials as local densifications in the lattice, it is not surprising that they may induce $\alpha$-like regions that are energetically more accessible than removal of an atom from the lattice. 

As a matter of fact, the electronic structure of the self-interstitial in $\delta$-Pu is quite reminiscent of the $\alpha$-Pu phase in that it induces large variations in spin polarization in its neighborhood. For the octahedral interstitial specie, the self-interstitial atom itself does possess a rather small spin moment of only $1.4~\mu_B$ within SP-GGA, and even a lower value of $1~\mu_B$ within SP+SO+OP-GGA . Its nearest-neighbor Pu atoms snap back to being $\delta$-like with large spin/orbital moments albeit with slightly reduced magnitudes. As was shown in Tab.~\ref{tab:f_dev}, large variations in magnitudes of localized spin/orbital moments between neighboring sites is a hallmark of the electronic structure of the $\alpha$-Pu phase. 

Hence the interstitial atom increases $f$-electron bonding in its immediate neighborhood in $\delta$-Pu. The spin polarization of the f-shell, which increases the occupation of the anti-bonding $f$-orbitals can be quantified by the FSM measure $\Delta \Sigma_F$, defined in Eq.~\ref{eq:fmom}. For the self-interstitial, $\Delta \Sigma_F^I = 0$, if it were to have NO effect on the $f$-electron bonding character of the host lattice. However, as is shown in Tab.~\ref{r-tab:deltamom}, $\Delta \Sigma_F^I$ are significantly less than 0, which is indicative of the increased bonding induced by the self-interstitial in $\delta$-Pu. They range from -1.15 calculated within (001)-polarized-SP+SO+OP-GGA to -0.65 within SP-GGA. These values correlated well with the self-interstitial relaxation volumes ranging from 0.85 within (001)-polarized SP+SO+OP-GGA to 1.32 within SP-GGA. Examining the cumulative FSM distributions $\Delta \Sigma_F^C(d)$, see Eq.~\ref{eq:dFSMc}, as exhibited in Fig.~\ref{fig:3} reveals that the three theories fully agree on the induced spin moment on the octahedral interstitial atom, but that the SP+SO+OP-GGA approximations relative to SP-GGA predict larger reductions of the spin moments on the first-neighbor shell but yield smaller effect on $f$-electron correlations beyond. Further comparison of Fig.~\ref{fig:3} with the cumulative formation volume distribution $\Delta V_F^C$ in Fig.~\ref{fig:3.5} unravels a strong correlation between the two distributions, as also was found in the case of the vacancy in Sect.~\ref{sec:vacancydelta}, We thus conclude that self-interstitials in $\delta$-Pu do cause local changes in $f$-electron  bonding, which strongly couple to the lattice deformations and result in much better accommodation of the interstitial defect in the $\delta$-Pu lattice compared to standard metallic systems such as Cu. 

Finally we discuss the formation compressibilities of the octahedral self-interstitial in $\delta$-Pu, as well as the (001)-dumbbell in fcc Cu. It can be seen that for the self-interstitials in $\delta$-Pu the two contributions to the formation compressibilities $\beta_0$ defined in Eq.~\ref{eq:beta0}, and $\beta_L$ defined in Eq.~\ref{eq:betaL} are comparable in size. This is in contrast to the case of the vacancy in $\delta$-Pu, in which $\beta_0 >> \beta_L$. As a result, the self-interstitial in $\delta$-Pu is much less compressible than the vacancy. Nevertheless, $\beta_0$ and $\beta_L$ for the (001)-dumbbell self-interstitial in Cu are much smaller in magnitude than the corresponding values for the octahedral interstitial in $\delta$-Pu. This is mostly due to the Cu bulk modulus being more than three times larger than that of $\delta$-Pu.

\bigskip

\subsubsection{Vacancies in $\alpha$-Pu}
\label{sec:vacancyalpha}

There are 8 crystallographically distinct lattice sites for vacancy and interstitial defects in $\alpha$-Pu. Table~\ref{tab:Adef_vac} gives the vacancy formation enthalpies and relaxation volumes versus crystallographic site. These calculations were conducted in 127-atom supercells with $2\times 2\times 2$ K-point grid, and full structural as well as supercell-shape relaxations were pursued. Furthermore, a correction for the 7\% overbinding error of the SP-GGA approximation is employed by applying an external mechanical tension at -5~GPa. The Bravais lattice vectors of the supercell $\vec{a}_i^S$ are obtained from the primitive 16-atom $\alpha$-Pu Bravais lattice vectors $\vec{a}_i^P$ as follows
\begin{eqnarray}
  \vec{a}_1^S &=& \vec{a}_1^P + 2\vec{a}_2^P, \nonumber\\
  \vec{a}_2^S &=& \vec{a}_1^P + 2\vec{a}_3^P, \\
  \vec{a}_3^S &=& \vec{a}_1^P - 2\vec{a}_3^P.\nonumber
\end{eqnarray}

In Tab.~\ref{tab:Adef_vac}, the vacancy formation enthalpies vary strongly between values as low as 0.31 eV (site 4) to as high as 1.33 eV (site 8). This is a typical feature of the $\alpha$-Pu structure, with its different inequivalent sites having vastly different Voronoi volumes, see Tab.~\ref{tab:A_vor_dev}, as well as different degrees of $f$-electron correlation represented by formation of localized spin/orbital moments, see Tab.~\ref{tab:f_dev}. It has been shown in the past \cite{GaStable:SADIGH:2005} that Ga impurities in $\alpha$-Pu prefer to substitute the Pu atom on site 8, which was shown to have important ramifications for the kinetics of $\delta$-to-$\alpha$ martensitic transformations in Pu. Note that site 8 has the largest Voronoi volume, see Tab.~\ref{tab:A_vor_dev}, and within SP-GGA is energetically the least favored site for the vacancy, see Tab.~\ref{tab:Adef_vac}. 

Based on said variations in the local atomic volumes of $\alpha$-Pu, one might have guessed that site 1, with the smallest Voronoi volume would be the most preferable vacancy site in this phase. However, as can be seen in Tab.~\ref{tab:Adef_vac}, sites 4 and 5 are by far the most favorable energetically. This implies that there are other interactions in $\alpha$-Pu than purely steric ones. For instance, using the FSM measure introduced in Eq.~\ref{eq:fmom}, it can be shown that vacancies induce increased spin polarization in $\alpha$-Pu. Table~\ref{tab:Adef_vac} reports the FSM values for each inequivalent vacancy in $\alpha$-Pu. It is noteworthy that except for site 8, all other vacancies possess FSM values that are positive and that are an order of magnitude larger than in $\delta$-Pu. Hence introduction of vacancies in $\alpha$-Pu induces spin polarization in this system, which in turn expands the lattice and thus compensates for the contraction that is conventionally expected to be induced by the vacant site. As a result, the relaxation volumes of all but the site-8 vacancy are nearly zero in $\alpha$-Pu. The anomalous behavior of this vacancy can be understood by noting the strongly negative FSM value of -3.3 associated with removal of an atom from site 8, which in turn indicates increased occupation of bonding orbitals, leading to contraction and thus a relaxation volume of -0.63. It is worth reminding the reader that this result should have been expected based on the large size of the Voronoi volume associated with site 8 in pure $\alpha$-Pu, which not surprisingly leads to formation of a large localized spin moment there, see Tabs.~\ref{tab:A_vor_dev} and \ref{tab:f_dev}.

\begin{figure}
\includegraphics[width=0.95\columnwidth]{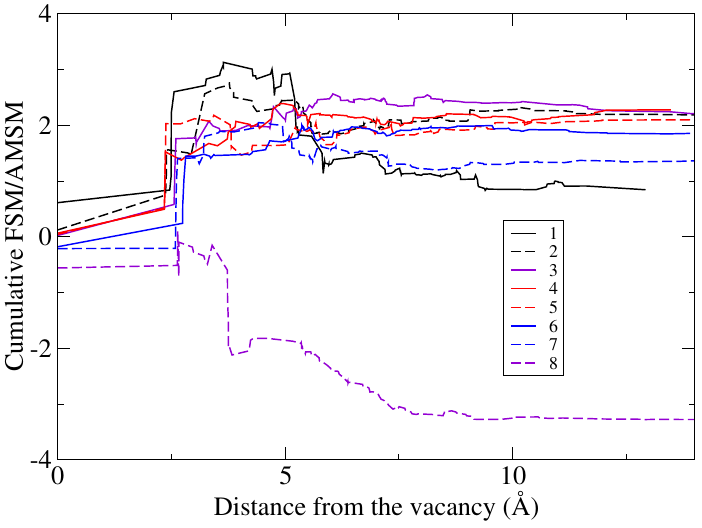}
\caption{Cumulative formation spin moment $\Delta \Sigma_F^C(d)$ as a function of distance $d$ from a vacant site in $\alpha$-Pu. It is the induced spin moment accumulated within a distance $d$ from a vacancy in a defect supercell. It is given in units of the average magnitude of per-atom spin moments (AMSM) in $\alpha$-Pu, calculated using the dilated SP-GGA (at -5~GPa) method applied to the eight inequivalent vacancies in $\alpha$-Pu. }
\label{fig:4}
\end{figure}

It is interesting to examine the cumulative FSM distributions $\Delta \Sigma_F^C(d)$ for the different vacant sites in $\alpha$-Pu. Due to the presence of eight inequivalent sites in this lattice, we have adopted the following definition for all distances $d > 0$ 
\begin{equation}
  \label{eq:aFSMc}
  \Delta \Sigma_F^C(d) = \sum_{i=1}^{N-1} \left(\sigma_i^d-\Sigma_L^{\kappa(i)}\right)~H(d-d_i).
\end{equation}
Above $\sigma_i^d$ is the local spin moment magnitude of the $i$th atom in the defect supercell, and $\Sigma_L^{\kappa(i)}$ is the spin moment of site $\kappa(i)$ in the perfect lattice, and $H(x)$ is the Heavyside function. Hence $\kappa(i)$ is a mapping of atom $i$ in the defect supercell to site $\kappa(i)$ in the perfect lattice. This is a reasonable definition for studying defect-induced spin moments in $\alpha$-Pu since the different sites in the perfect lattice are so dramatically different from each other. In order for $\Delta \Sigma_F^C(d)$ in Eq.~\ref{eq:aFSMc} to asymptotically approach the thermodynamic FSM value $\Sigma_F$ at large distances $d$, we need to define
\begin{equation}
  \label{eq:aFSMc0}
  \Delta \Sigma_F^C(0) = \bar{\Sigma}_L - \Sigma_L^v,
\end{equation}
where $\bar{\Sigma}_L$ is the average magnitude of the per-atom spin moments (AMSM) in $\alpha$-Pu, and $\Sigma_L^v$ is the magnitude of the spin moment of the vacant site in defect-free $\alpha$-Pu. Figure~\ref{fig:4} shows the spatial distributions of $\Delta \Sigma_F^C(d)$ for all the eight inequivalent vacancies within SP-GGA. It can be seen that except for the eighth site, the Pu atoms within a 5~\AA ~shell around the vacancy increase their spin moment sizes, leading to a positive $\Delta \Sigma_F^C$ at $d<5$~\AA. All FSM values, including that of the eighth site converge within 7.5~\AA. This is quite different from the vacancy in $\delta$-Pu. see Fig.~\ref{fig:2}, which exhibits a long-range distribution of volume as well as spin moment relaxations. We thus conclude that in $\alpha$-Pu, coupling of the defect-induced changes to the $f$-electron correlations are quite short range. In order to clearly illustrate the relation between the formation spin moments and volume relaxations, we calculate the properties of the point defects in $\alpha$-Pu within non-magnetic GGA at 0~GPa, which corresponds to an average atomic volume of 17.6~\AA $^3$. The results are listed in in Tab.~\ref{tab:alphanm}. It is clear that in the absence of spin polarization the relaxation volumes are reduced and range from -0.2 to -0.5 atomic volumes. 

One source of concern about the accuracy of the dilated (-5~GPa) SP-GGA approximation that is used here to calculate properties of point defects in $\alpha$-Pu is the very low formation enthalpy found for the site-4 vacancy in Tab.~\ref{tab:Adef_vac}. If correct, it implies a very high thermal equilibrium vacancy concentration. While equilibrium point defect concentrations do not enter calculations of void swelling bias within the classic theory of damage-induced void growth in materials, it is quite likely to significantly delay the incubation times for void nucleation. However, it is not inconceivable that dilated SP-GGA may appreciably underestimate vacancy formation enthalpies for several of the sites in $\alpha$-Pu. The reason for this assessment mainly falls back on the strong heterogeneous nature of the $\alpha$-Pu structure. It is hard to see that the effect of SO+OP in this system can be simply replaced by a homogenous $PV$ term, as described in Sect.~\ref{sec:mod-sp}, which has been quite successfully applied to the study of point defects in $\delta$-Pu earlier in this paper. It is in fact quite easy to argue that -5~GPa tension can cause large reduction of the vacancy energies in $\alpha$-Pu without much effect on the formation volumes. The reason for this is that the vanishing vacancy relaxation volumes in $\alpha$-Pu imply $\Delta V_F \sim 1$ atomic volume, and the latter is the pressure-derivative of the formation enthalpy, see Eq.~\ref{eq:relV}. which implies that vacancy formation enthalpies should be about 0.6 eV higher, if calculated at 0~GPa. This argument is supported by the non-magnetic GGA defect results in Tab.~\ref{tab:alphanm}, showing formation enthalpies ranging from 1.2 eV 2.1 eV. It is interesting to note that at the non-magnetic equilibrium volume of 17.6 \AA $^3$, site 8 and site 1 are the lowest energy vacancy sites. Curiously, site 8 is the least energetically favored site for the vacancy at 19.3 \AA $^3$. see Tab.~\ref{tab:Adef_vac}. However, it has by far the smallest formation volume, which leads to its relative stabilization compared to the other vacancy species as pressure is increased from -5 GPa to 0~GPa. On the other hand, pressure-derivative of the relaxation volume measured in units of atomic volume is determined by $\beta_0$, see Eq.~\ref{eq:beta0}. Due to the relatively stiff $\alpha$-Pu lattice, compared to $\delta$-Pu, one may reasonably expect that $\beta_0$ does not exceed 0.01~(GPa)$^{-1}$ in $\alpha$-Pu. This implies that the maximum change expected for  $\Delta V_R$ measured in units of atomic volume over 5 GPa is less than 0.05. Hence, while the formation enthalpies can be strongly dependent on dilatation pressure in $\alpha$-Pu, the formation volumes are quite insensitive to it. As a result, the SP+SO+OP-GGA approximation may be necessary for obtaining accurate estimates of enthalpies of formation of point defects in $\alpha$-Pu. We defer this study to future works.

Nevertheless, the three qualitative results found in this section with the collinear SP-GGA theory will stand: (i) the magnitudes of the vacancy relaxation volumes in $\alpha$-Pu are an order of magnitude smaller than those of $\delta$-Pu and their distribution is much more short range, (ii) the equilibrium vacancy concentrations in $\alpha$-Pu at finite temperatures are higher than in $\delta$-Pu, and (iii) the two sites with the lowest vacancy enthalpies are 4 and 5, which together with their periodic replicas inscribe two-dimensional planes spaced over 9 {\AA} apart. This heterogeneous spatial distribution can have important implications for vacancy diffusion in the material, although activation barrier heights will have to be calculated before conclusive predictions can be made.

\bigskip

\begin{table}{}
  \caption{Vacancy properties in 128-atom supercells of $\alpha$-Pu, calculated within SP-GGA under dilational stress conditions, $p_0 =$ -50.0  kbar;  $\Omega_{at}=19.3$\AA$^3$ is the atomic volume for the bulk. The average magnitude of the atomic spin moment in the bulk is $2.66~\mu_B$.}
    
\centering
\begin{tabular}{c |c c c }
\hline
 Site &     Formation         & Relaxation vol.      &Formation mom.       \\
        &      enthalpy (eV)   &  / atomic vol.         & / atomic mom.          \\ [1ex]
\hline
  1    &    1.02                   & -0.02                     &   0.9   \\
  2    &     0.64                  & -0.05                     &   2.2   \\
  3    &     0.87                  &  0.06                     &   2.2   \\
  4    &     0.31                  & -0.14                     &   2.3   \\
  5    &     0.47                  & -0.07                     &   2.1   \\
  6    &     0.89                  &  0.05                     &   1.9   \\
  7    &     0.98                  & -0.01                     &   1.4   \\
  8    &     1.33                  & -0.63                     &  -3.3   \\[1ex]
\hline

\end{tabular}
\label{tab:Adef_vac}
\end{table}

\begin{table}{}
\caption{Interstitial properties in 128-atom supercells of $\alpha$-Pu, calculated within SP-GGA under dilational stress conditions, $p_0 =$ -50.0  kbar;  $\Omega_{at}=19.3$\AA$^3$ is the atomic volume for the bulk. The average magnitude of the atomic spin moment in the bulk is $2.66~\mu_B$. 
}
\centering
\begin{tabular}{c |c c c }
\hline
 Site &     Formation         & Relaxation vol.      &Formation mom.            \\
        &      enthalpy (eV)   &  / atomic vol.         & / atomic mom.             \\ [1ex]
\hline
  1    &   1.66                     & 2.47   & 5.3 \\
  2    &   2.07                     & 2.22   & 4.9 \\
  3    &   1.19                     & 2.24   & 3.5 \\
  4    &   1.70                     & 2.16   & 2.7 \\
  5    &   2.10                     & 2.42   & 3.0 \\
  6    &   2.12                     & 2.33   & 3.9 \\[1ex]
\hline

\end{tabular}
\label{tab:Adef_int}
\end{table}

\begin{table}{}
  \caption{The table lists the formation energies and relaxation volumes of all the inequivalent vacancies and distinct self-interstitial defects found in this study in $\alpha$-Pu. The calculations are conducted within within non-magnetic GGA at 0~GPa. In this approximation, the calculated atomic volume of the bulk $\alpha$-Pu lattice is $\Omega_{at} = 17.6$~\AA$^3$. }
  \centering
\begin{tabular}{c |c  c | c  c}
\hline
Site&\multicolumn{2}{c|}{Vacancy}&\multicolumn{2}{c}{Self-interstitial}\\
\hline
i &  Form. Energy  & Rel. Vol. & Form. Energy & Rel. Vol.\\
  &  (eV)          &  (at. vol.) & (eV) & (at. vol.) \\ [1ex]
\hline
  1     &   1.2   &   -0.23       &   1.2    &  1.1  \\
  2     &   1.8   &   -0.42       &   1.6    &  1.2  \\                
  3     &   1.6   &   -0.35       &   1.7    &   1.4  \\        
  4     &   1.4   &   -0.45       &   2.5    &  1.5   \\              
  5     &   1.6   &   -0.35       &   1.1    &  1.3   \\              
  6     &   2.1   &   -0.21       &   2.0    &  1.5   \\                 
  7     &   2.1   &   -0.24       && \\                 
  8     &   1.2   &   -0.23       &&  \\ [1ex]                 
\hline

\end{tabular}
\label{tab:alphanm}
\end{table}

\subsubsection{Self-interstitials in $\alpha$-Pu}

\begin{figure}
\includegraphics[width=0.95\columnwidth]{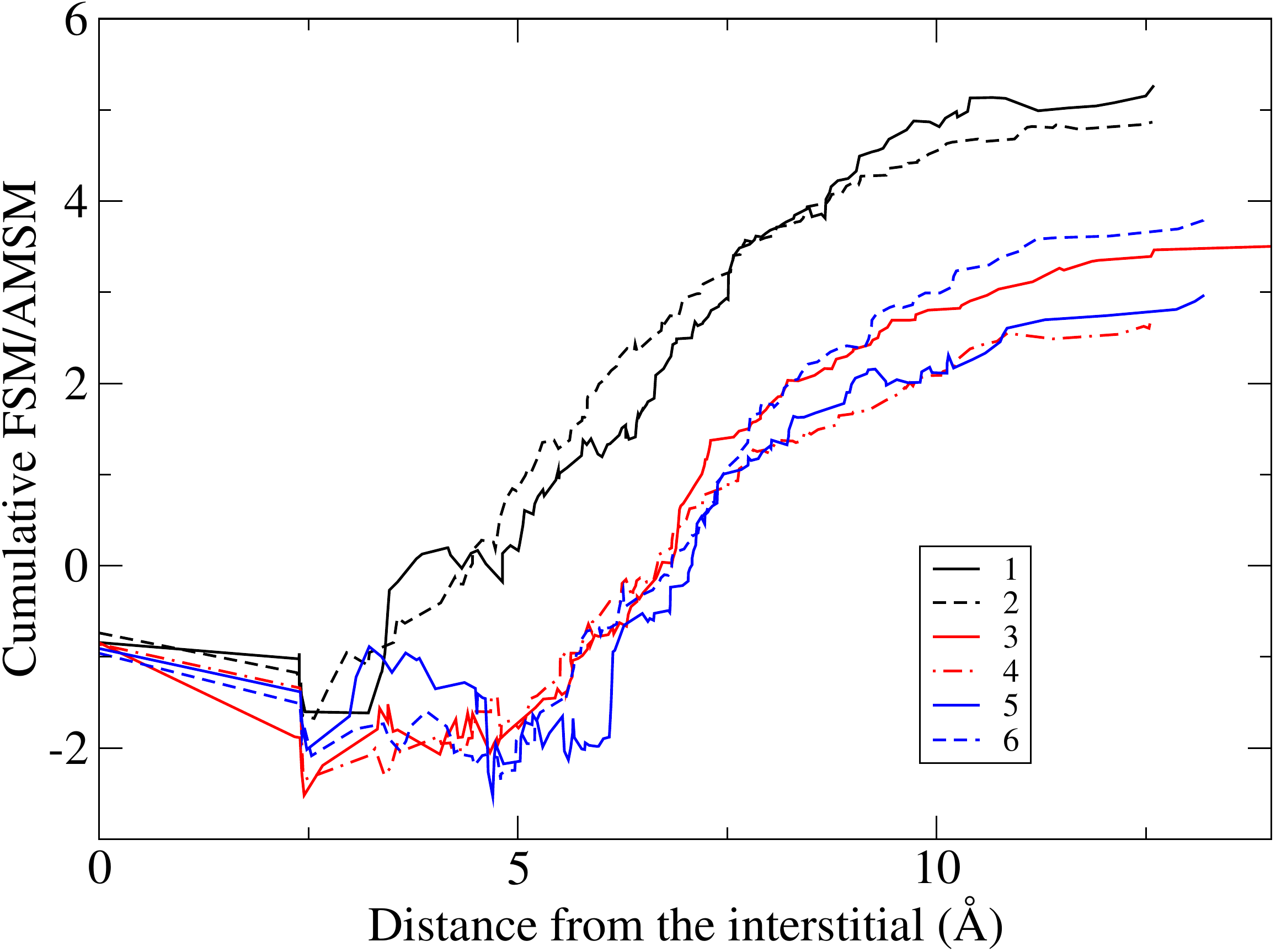}
\caption{Cumulative formation spin moment $\Delta \Sigma_F^C(d)$ as a function of distance $d$ from a self-interstitial site in $\alpha$-Pu. It is the induced spin moment accumulated within a distance $d$ from a self-interstitial in a defect supercell. It is given in units of the average magnitude of per-atom spin moments (AMSM) in $\alpha$-Pu, calculated using the dilated SP-GGA (at -5~GPa) method applied to the six distinct self-interstitial defects found in $\alpha$-Pu. }
\label{fig:5}
\end{figure}

The self-interstitial calculations in $\alpha$-Pu have been performed using the same electronic and supercell structure parameters as for the vacancies in this phase. However, the positions of low-energy self-interstitials in this low-symmetry phase were unknown. Hence we chose starting guesses for candidate self-interstitial defect positions by searching for centers of the largest empty holes in the lattice. They correspond to Voronoi vertices belonging to the largest Voronoi cells in the lattice. For this purpose, we have followed a sequential process for finding new candidate positions:
\begin{itemize}
\item Find the Voronoi vertex in the lattice, about which the largest sphere can be drawn that contain no lattice points.
\item Add this center and all its symmetry-equivalent points to the lattice.
\item Repeat.
\end{itemize}

In this way, we have identified six distinct self-interstitial positions in the $\alpha$-Pu structure. Table~\ref{tab:Adef_int} reports their formation enthalpies and relaxation volumes after complete structural relaxations including cell-shape changes have been conducted. The calculated formation enthalpies span values from 1.19~eV to 2.12~eV. While these values are much higher than the self-interstitial formation enthalpies in $\delta$-Pu, they are quite small relative to normal close-packed metals. For example, the calculated value for Cu within DFT-PBE is 3.0~eV, see Tab.~\ref{r-tab:deltavac}. In spite of the rather small formation enthalpies in $\alpha$-Pu, the corresponding relaxation volumes are quite large, spanning values from 2.16~at. vol. to 2.47~at. vol., which should be compared to 1.9~at. vol. for Cu. This is at first sight a surprising result. In normal materials, a clear positive correlation exists between formation energies of lattice defects and magnitudes of deformation they introduce in their hosts. The apparent disconnect in $\alpha$-Pu between formation energy and relaxation volume can be traced to $f$-electron spin moments induced by addition of a self-interstitial atom in the lattice. For this purpose, the FSM values for the different self-interstitial species in $\alpha$-Pu are listed in Tab.~\ref{tab:Adef_int}. The values span a range from 2.7~AMSM to 5.3~AMSM, by far largest than any other defects encountered in this paper. Hence, promotion of $f$-electron spin polarization by the self-interstitials in $\alpha$-Pu leads to large relaxation volumes. It should be noted that for the purpose of calculation of void swelling bias, the relaxation strain is of primary interest.

In order to better understand the coupling of $f$-electron correlations to self-interstitials in $\alpha$-Pu, we study the distributions of cumulative FSM $\Delta \Sigma_F^C(d)$, as defined in Eq.~\ref{eq:aFSMc} for $d > 0$, and as follows for $d=0$
\begin{equation}
  \label{eq:aiFSMc0}
  \Delta \Sigma_F^C(0) = \sigma_{int}^d - \bar{\Sigma}_L.
\end{equation}
In the above equation, $\sigma_{int}^d$ is the spin moment magnitude of the interstitial atom, and $\bar{\Sigma}_L$ is the average magnitude of the per-atom spin moments (AMSM) of $\alpha$-Pu. Figure~\ref{fig:5} depicts the distributions of $\Delta \Sigma_F^C(d)$ for the six self-interstitial species found in $\alpha$-Pu. It appears that the interstitial atom and its immediate neighbors undergo compression and reduction of their spin moment magnitudes, while beyond a radius that can be either $\sim 3$~\AA ~ or $\sim 5$~\AA , depending on the self-interstitial site, dramatic increase in spin moments followed by volume expansion takes place. It is clear that both the spin-density response as well as the tensile strain field emanating from self-interstitials in $\alpha$-Pu are quite long-range and likely require larger supercells than 129-atom ones used in this study to reach convergence. In order to verify the effect of induced spin moments on lattice expansion due to self-interstitials, we compare with relaxation volumes calculated within non-magnetic GGA at 0 GPa, see Tab.~\ref{tab:alphanm}. Under these conditions, the relaxation volumes are much reduced, quite as expected. They range from 1.1 to 1.5 atomic volumes. Hence, the coupling of the self-interstitials to $f$-electron correlations in $\alpha$-Pu is crucial to the large relaxation volumes predicted for these defects in this system. 

Finally, examining the two lowest-energy self-interstitial sites in Tab.~\ref{tab:Adef_int}, i.e. 3 and 4, it is found that they comprise layers of atoms spaced over 7 {\AA} apart.  The next most favored interstitial site lies significantly (0.5 eV) higher.
This suggests that self-interstitial diffusion may also be quasi-two-dimensional, although again the transition state barriers need to be calculated.
In particular, transport in two-dimensions is not well-described by mean field approximations.
This suggests that the conventional rate theory of void swelling may be complicated in $\alpha$-Pu.
Both vacancies and interstitials favor site 4;  this may increase the cross section for annihilation over conventional expectations based on uniform distributions and isotropic diffusion.

\section{Discussion} 
\label{sect:discussion}
\subsection{Aging and Void Swelling}
To date, comparatively little research has been reported on radiation-induced aging in $\alpha$-Pu, even though it is the thermodynamic equilibrium phase at ambient conditions.
In contrast, many aspects of the problem  have been studied in metastable, alloy-stabilized  $\delta$-Pu including compositional changes due to decay/transmutation, changes in density\cite{
density:2000, CHUNG:2006, CHUNG:2006, CHUNG:2016}, collision cascades
 \cite{KUBOTA:2007, BERLU:2008a, BERLU:2008b}, evolution of damage and defect populations \cite{resistivity:FLUSS:2004, magDefects:BACLET2002, JOMARD:2007, magDefects:MCCALL:2006, ANISIMOV:2013, CONRADSOND:2014a, KARAVAEV:2016, deltaDefects:HERNANDEZ:2017, magDefects:HERNANDEZ2020}, including He bubbles\cite{bubble:SCHWARTZ:2005,DREMOV:2008,JEFFRIES:2011,AO:2012,KARAVAEV:2014,ROBINSON:2014,WU:2017}, changes in bulk properties 
\cite{VALONE:2007, DREMOV:2013, DREMOV:2011, ARSENLIS:2005, LI:2011, KARAVAEV:2017, puAging:Modulus:Exp}, and the prospect of void swelling\cite{ALLEN:2015a,ALLEN:2015b}.

So far, there have been no experimental reports of void swelling in $\delta$-Pu, despite having samples irradiated over decades and despite calibrated attempts at 
accelerated aging in more recent years. It has been proposed that the so-called net bias factor may be negligible in $\delta$ in which case void swelling should be insignificant \cite{ALLEN:2015a}. The bias is the crux of the phenomenon:\cite{WOLFER:ASHKIN:SIPA:1975a,WOLFER:ASHKIN:SIPA:1975b,SNIE} there would be no void growth and no swelling in the absence of biased diffusion and defect segregation. However, this expectation is predicated on models of elastic coupling between stress fields from the microstructure and mobile defects using parameters taken from {\it ab initio} DFT calculations. It is unclear if this conventional approach can be directly extended to as complex a metal as Pu. Such uncertainty makes the behavior of other phases of Pu of  interest. The conventional theory (including the DFT-derived parameters) would gain support if it can successfully explain swelling behavior in a diversity of Pu phases. Arguably, finding discrepancies with experiment would be even more informative.

In practice, the conventional theory explains both peak rates and trends in void swelling in ordinary metals, but the approximations made in the analysis and the uncertainties in the underlying parameters make the results as much explanatory or extrapolative as predictive. Environmental parameters like temperature, pressure, and radiation flux (or self-decay rates in Pu)  may be well-known for a given system. However, the microscopic response that immediately follows radiation bombardment (effectively the average number of atoms displaced and the average number of point defects ultimately left behind by a single Pu-decay) is uncertain. Typically, reasonable parameter choices can be made that quantitatively reproduce experimental swelling behavior  in, e.g., reactor structural materials. Theory-based modeling can then explain a range of key swelling properties \cite{review:WOLFER:2000}, viz:
1) There is an incubation period and transient initial delay during which swelling and void growth is absent or greatly reduced. 
2) Once the peak swelling rate (measured against the rate of radiation-driven atomic displacements) is achieved it can continue fairly steady-state even as the relative volume changes exceed 100\%. 
3) The peak swelling rate for a given material is also fairly independent of temperatures and radiation fluxes over a wide range, as well as initial microstructure.
4) In contrast,  the incubation delay can vary greatly depending on the starting microstructure, temperature, and radiation dose rate.

That final point is understood from classical nucleation theory: under a vacancy supersaturation, voids typically nucleate heterogeneously at pre-existing helium bubbles, thus void swelling cannot commence until bubbles have grown to the critical size \cite{SURH}. A high concentration of defect traps like small precipitates can delay the onset of void swelling by absorbing or binding mobile defects like helium, vacancies, and interstitials. This leads to many, smaller bubbles (delaying the first bubbles from reaching critical-size), and it also facilitates the annihilation of intrinsic defects by trapping both vacancies and interstitials together. However, such extrinsic, microstructure-dependent delays mean that a null swelling result from a given experiment does not preclude its appearance at later times. Indeed, the microstructure is expected to change substantially during prolonged irradiation. Under long exposures, each atom in a material can be displaced from its lattice site on average 10 or 100 times, usually settling at a completely new lattice site and only occasionally forming a point defect or defect cluster. In light of this prospect, the theoretical intrinsic net bias towards swelling may provide valuable insight into possible long-term void swelling behavior \cite{SNIE}.

In the theory, steady radiation bombardment generates quasi-stationary vacancy and interstitial populations such that defect creation balances loss by mutual annihilation or by absorption at microstructural sinks in the system. The relevant microstructure includes network dislocations, dislocation loops, cavities (high gas-density helium bubbles and low density voids), precipitates, grain boundaries, etc. In many analyses, stochastic defect production and spatially-dependent fluxes within a real microstructure are approximated by mean field, steady-state diffusion calculations in idealized geometries. The microstructure evolution may be comparably simplified: cavities evolve according to mean field reaction rates while dislocations multiply and annihilate according to simple rate equations for the aggregate density \cite{SURH}. Often the mean field problem is simplified further by assuming isotropic diffusivity and by taking the biasing interactions to be the product of relaxation volumes of mobile defects and hydrostatic stress fields of microstructure sinks.

The computed average flux of vacancies and interstitials to each type of sink is then summarized by separate bias factors, each a single dimensionless number comparing the diffusive flux in the presence of the mutual interaction and in the absence (i.e., unbiased diffusion). Their relative values determine the overall tendency to vacancy-interstitial segregation and swelling. Network dislocations contribute the most significant long-range elastic stresses, so their bias factors dominate the situation. E.g., dislocations of density $\rho$ (m$^{-2}$) are modeled as an infinite straight edge dislocation in a cylindrically-symmetric region with a 2-D Wigner-Seitz radius, where $\pi R_{dis}^2 =1/\rho$. Solving for the diffusion with and without an elastic coupling that decays with distance as $1/r$, the dislocation bias factor is approximated by:
\begin{eqnarray}
  Z_{dis}(T,\Delta V_R) &&=  ln\biggl(\frac{R_{dis}}{r_{dis}^{cor}}\biggr) \times \nonumber \\
 && \bigg\lbrack\frac{K_0(r_{dis}^{cap}/R_{dis})}{I_0(r_{dis}^{cap}/R_{dis})}-\frac{K_0(r_{dis}^{cap}/r_{dis}^{cor})}{I_0(r_{dis}^{cap}/r_{dis}^{cor})}\bigg\rbrack^{-1}
\end{eqnarray}
in terms of modified Bessel functions, where $\Delta V_R$ is the relaxation volume (Eq.~\ref{eq:relV1}) of the mobile point defect, $R_{dis}$ is the 2D Wigner-Seitz radius, $r_{dis}^{cor} = 2b$ is the dislocation core radius (taken to be twice the burgers vector, $b$), and $r_{dis}^{cap}=\frac{(1+\nu)^2 }{36 \pi (1-\nu)} \frac{\mu |\Delta V_R|}{k_B T}$ is the capture radius at which the point defect is absorbed by the sink in terms of shear modulus, $\mu$, and Poisson ratio, $\nu$. Voids, the other important class of sinks in this problem, have comparatively weak stress fields and become increasingly unbiased sinks, $Z_{cav}\simeq1$, in the limit of large sizes.

The net difference at which interstitials and vacancies are absorbed then determines whether dislocations climb and lattice sites are added while cavity volume increases; this is expressed by the {\it net} bias factor. If the predicted value is small, there should be little impetus for void swelling, incubation periods can be prolonged, and any eventual swelling will be gradual. Assuming that the overall total sink strength is dominated by the dislocation contribution \cite{SNIE}, the net bias factor becomes: 
\begin{equation}
B \simeq \frac{Z_{dis}^{int}}{Z_{dis}^{vac}} - \frac{Z_{cav}^{int}(x)}{Z_{cav}^{vac}(x)}
\label{netBias}
\end{equation}
as a function of cavity size, $x$. The most favorable conditions for void growth are obtained when $Z_{cav}\rightarrow1$, which will be assumed here. The resulting material property, $B$, can be estimated from first principles calculations.

However because of the assumed isotropy in the derivation of Eq.~\ref{netBias}, the result is only applicable to materials like cubic $\delta$-Pu. The site-dependent defect energies for $\alpha$-Pu in Tabs.~\ref{tab:Adef_vac} and~\ref{tab:Adef_int} already suggest that diffusion will be highly anisotropic, potentially affecting the calculations for net bias. The low crystal symmetry in $\alpha$ also implies anisotropic lattice relaxation around defects, which requires coupling to the full elastic stress tensor. Nevertheless, the relative V and SI relaxation volumes are still primary quantities of interest here, and so we simply compare relaxation volumes between $\alpha$- and $\delta$-Pu as surrogates for the relative tendency to void swelling. Based on Eq.~\ref{netBias} and the computed defect relaxation volumes, swelling would be weak or nonexistent in $\delta$-Pu but based on the defect volumes in $\alpha$-Pu it is significantly likelier to be seen there.

\subsubsection{$\delta$-Pu}

Despite the simple crystal structure, there are significant uncertainties for void swelling properties in $\delta$-Pu. Given the well-known errors for equilibrium volumes of Pu phases, standard approaches like SP-GGA could have difficulty with the defect relaxation volumes. Besides the usual elastic interactions, recent calculations in (static mean field) SP-GGA also find a magnetic component to Pu defects \cite{SHORIKOV:2013,spinDelta:HERNANDEZ:2019,magDefects:HERNANDEZ2020}, which would contribute to the total interactions and bias factors. There may furthermore be low energy electronic/magnetic excitations of the defects with significant effects on the relaxation volume, as considered in this work. Table~\ref{r-tab:flips} includes the SP-GGA results for a number of metastable magnetic excitations in the 108-site $\delta$-Pu vacancy supercell. The tabulated configurations represent a fraction of all possible nearest-neighbor site combinations, but they are enough to show that magnetism may be significantly coupled to lattice degrees of freedom. However as yet, there is no evidence of a similar magnetoelastic effect for the interstitial in $\delta$-Pu near the experimental density. This asymmetry may be intrinsic or it may be an artifact of the external tension applied in the SP-GGA calculations to approximate the experimental density. The simple vacancy spin-flips in Table~\ref{r-tab:flips} are all disfavored in the more accurate SP+SO+OP-GGA approximation. Non-collinear excitations may occur instead.

Independently of any particular DFT approximation, electronic fluctuations can influence the defect volume whenever the energy for an atom to adjust its electronic state (and f-shell size) is less than the energy stored in elastic strain. I.e. such a localized excitation, $\Delta E$, could become favorable if it alters an atomic volume by $\Delta V$, where the local pressure gives ${-P \Delta V}/{\Delta E} >1$. Some similar dimensionless parameter argument may relate to the unusual bulk polymorphism and thermal expansion properties as well. For example, thermally-accessible  electronic excitations have been suggested to explain the anomalous thermal expansion or Invar effect of $\delta$-Pu.

It is notable that all of the vacancy energies listed in Table~\ref{r-tab:flips} are thermally-accessible. A partition function for this limited state space can provide expectation values for {\it classical} thermal fluctuations: the thermal average vacancy relaxation volume ranges from -0.25 atomic volumes at 300K to -0.29 at 600K. The results suggest that the net bias factor could be temperature-dependent in $\delta$-Pu. However, the variation is only of order 10\%, probably not enough to affect accelerated aging experiments. Some temperature dependence is not unprecedented in void swelling; here it is a magnetostructural version of the paraelastic effect already seen in some other materials.

The large volume range for the $\delta$-Pu vacancy also implies that the thermal-average relaxation volume will further depend on the local hydrostatic stress. Similar so-called `diaelastic' defect effects (also called modulus interactions) are also seen in some systems. It means that the expected vacancy relaxation volume will be weakly position dependent, affecting the mean field diffusion analysis for the bias factors. For completeness, Table~\ref{r-tab:deltavac} includes estimates for the effective compressibilities of the ground-state vacancy and self-interstitial species. In particular, the large effective compressibility found for the vacancy implies that the modulus effects can be quite significant in plutonium. 


The different DFT approximations predict a considerable range of defect parameters, but all of them imply that void swelling is weak or nonexistent in $\delta$-Pu. 
The most extreme case is for the SP-GGA results at the theoretical equilibrium; here the distorted monoclinic vacancy has a larger absolute relaxation volume than the interstitial. In that case, the net bias, $\Delta B$ is actually negative; no swelling is then expected as the vacancies would be preferentially drawn to dislocations.
If the SP+SO+OP-GGA results are used, v=-0.9, i=0.85, and the net bias is effectively zero. The SP-GGA results at -3GPa external stress depend on which distribution of magnetic vacancy states is considered. For the putative vacancy ground state (with the smallest relaxation volume), $\Delta B<0.2$; this net bias is comparable to that for ferritic steels, which do not swell much \cite{SNIE}. Thus, essentially no DFT parameter set for $\delta$-Pu predicts much void swelling (i.e., DFT predicts long incubation times and slow steady swelling, if any at all).

\subsubsection{$\alpha$-Pu}

We also examine defect formation energies and relaxation volumes in $\alpha$-Pu, a more complicated structure with 8 crystallographically-distinct sites. 
SP-GGA vacancy and interstitial calculations have been performed under -5 GPa hydrostatic tensile stress for all of the sites. This is sufficient for a preliminary comparison of void swelling tendencies with $\delta$-Pu. Interstitial formation energies are consistently lower than vacancies. While the lowest vacancy formation energy suggests an extremely high thermal equilibrium density, the other striking result is that the relaxation volumes are close to zero for almost all of the lattice sites, with large or small formation energies alike.  In contrast, the interstitial relaxation volumes are all quite large.  This would imply that $\alpha$-Pu is as prone to void swelling as aluminum (one of the faster swelling materials known), although the low symmetry and potentially anisotropic diffusion may affect this prediction. The incubation delay will also depend on impurities and the starting microstructure.

Remarkably, the two lowest energy sites are located in a layer-like arrangement for both interstitials and vacancies. Intrinsically anisotropic bulk diffusion in low-symmetry crystal structures may influence the mean field diffusion fields to voids versus dislocations and so affect the bias \cite{WOO:anisotropic:1988,BORODIN:1993}. Furthermore, if the diffusion approaches quasi two-dimensionality it could invalidate the mean field approximations for the bias factors. Conceivably, the spatial arrangement of surrounding sinks (bubbles or voids and dislocations) could then become important.

The lower energy defect sites also coincide for vacancies and interstitials, which may substantially increase their annihilation rates. In contrast, the lowest energy sites for Ga are well-separated from the preferred vacancy positions \cite{GaStable:SADIGH:2005}.  This may inhibit Ga diffusion and prolong the metastability of alloy-stabilized Pu.  Substitutional He may conceivably show similar behavior. In this light, a SP-GGA study of He defects is already planned.

\section{Conclusion}

In this paper, we report extensive first-principles calculations of point defect energies and structures in the $\alpha$- and $\delta$-Pu phases. 
Two broad surveys of point defect properties are reported in the SP-GGA approximation along with a separate study using the SP+SO+OP-GGA approximation,
a method that predicts accurate bulk phase behavior.
We discuss the likelihood of void swelling for these two materials in light of the calculated defect parameters.
As might be expected from the equilibrium phase diagram, defect behaviors are complex in this material, and this naturally complicates any analysis of swelling behavior.
However, conventional elastic interactions will remain relevant to the theory even if other, more novel mechanisms may contribute.
The calculated material properties suggest that void swelling is a greater possibility in $\alpha$-Pu.
Void swelling is expected to be weak or nonexistent in $\delta$-Pu, and indeed no void swelling has been reported from experiment.

\subsubsection{Acknowledgements}

We acknowledge Hector Lorenzana and Scott McCall for helpful discussions, and for making us aware of their recent experimental progress by private communication.  We also thank A. Arsenlis, N. Barton, and J. Belof for project support. This work was performed under the auspices of the U.S. Department of Energy by Lawrence Livermore National Laboratory under Contract DE-AC52-07NA27344.
 
\biblio

\appendix*
\subsection{Appendix: Finite supercell-size errors}

Our goal in this appendix is to study size-convergence of defect energies under supercell volume/shape constraint as compared to under applied external stress. For this purpose, consider a perfect crystal structure at pressure $P_0$, temperature $T_0$, and non-hydrostatic stress state $\sigma_0$. For brevity, we only consider here a single stress component, but generalization to a full stress tensor is straightforward. As an example of a crystal under a non-hydrostatic stress, let us consider an antiferromagnetic fcc crystal, composed of ferromagnetic (001) layers, with the adjacent (001) layers having opposite spin moments, i.e. L10 spin ordering. This is the lowest-energy spin configuration for $\delta$-Pu within the SP-GGA approximation, and happens to break the cubic symmetry. Hence, in general the cubic fcc structure will be under a non-zero stress state that couples to tetragonal shear distortion of the crystal leading to $c/a$-ratio that deviates from unity. Let us denote by $\eta_L$ the tetragonal strain state of the perfect crystal corresponding to the stress $\sigma_0$, and by $V_L$ its volume at pressure $P_0$.

We now define the canonical energy as a function of volume and strain $E(V,\eta)$, with the properties
\begin{eqnarray}
  \label{eq:sstrain}
  \frac{\partial E(V,\eta)}{\partial \eta} &=& \sigma.\\
  \frac{\partial E(V,\eta)}{\partial V} &=& -P.
  \label{eq:spres}
\end{eqnarray}
These equations state the relevant stress-strain relationships. Before proceeding, we define two relevant elastic constants 
\begin{eqnarray}
  \label{eq:elcnst}
  \frac{\partial^2 E(V,\eta)}{\partial \eta^2} &=& C^{\eta}.\\
  \frac{\partial^2 E(V,\eta)}{\partial V^2} &=& -\frac{K}{V}.
  \label{eq:bulk}
\end{eqnarray}
Note that $K$ is the bulk modulus, and in the current example, $C^{\eta}$ is the elastic constant for tetragonal shear. 

We now proceed to define the enthalpy by a Legendre transformation of the canonical energy 
\begin{eqnarray}
  H(P,\sigma) &=& E(V(P,\sigma),\eta(P,\sigma)) + P~V(P,\sigma) \nonumber\\
                  && - ~ \sigma~\eta(P,\sigma),
  \label{eq:Hsig}
\end{eqnarray}
where $\eta(P,\sigma)$ are $V(P,\sigma)$ are obtained by solving Eqs.~\ref{eq:sstrain} and ~\ref{eq:spres}. For the example of the perfect crystal described above, the different variables in Eq.~\ref{eq:Hsig} can be defined as follows: $P\equiv P_0$, $\sigma\equiv \sigma_0$, $V(P_0,\sigma_0)\equiv V_L$, and $\eta(P_0,\sigma_0)\equiv \eta_L$.

Now consider the energetics of this crystal upon introduction of a concentration $c_0$ of point defects. The energy and the enthalpy functions can now be generalized to incorporate a finite point-defect concentration, denoted by $\tilde{E}(V_L,\eta_L,c_0)$, whenever system's density and strain are constrained, and $\tilde{H}(P_0,\sigma_0,c_0)$, whenever the external pressure and stress are controlled. Of course in practice, the defect properties are calculated in periodic supercells at finite defect concentration. Hence finite difference formulas are used to approximate partial derivatives.

In the following, we consider two different procedures: (i) calculating formation energy $\left.\frac{\partial \tilde{E}}{\partial c}\right|_{V_L,\eta_L,c_0}$ of point defects allowing for no relaxation of either cell shape or volume, and (ii) calculating formation enthalpy $\left.\frac{\partial \tilde{H}}{\partial c}\right|_{P_0,\sigma_0,c_0}$ of point defects allowing for relaxations of both supercell shape and volume. In this case, we can also define a relaxation volume as in Eq.~\ref{eq:relV}, and analogously a relaxation strain
\begin{equation}
  \Delta \eta_R = \frac{\partial \tilde{H}}{\partial c}.
\end{equation}

Let us then start with case (i). In this case, the defect supercell volume $V_D$ as well as its strain state $\eta_D$ are kept fixed and equal to the perfect lattice $V_L$ and $\eta_L$, i.e. $V_D = V_L$, and $\eta_D = \eta_L$. The finite-size error of the formation energy calculated in a defect supercell containing N sites can be expanded in a Taylor serier, whose first term is
\begin{equation}
 \tilde{E}(V_L,\eta_L,1/N) - \tilde{E}(V_L,\eta_L,0)  - \Delta E_F = 
  \frac{c_0}{2}\frac{\partial^2 \tilde{E}}{\partial c^2}  + ... .
\end{equation}
Note that $\Delta E_F = \partial \tilde{E}/\partial c$. 

Now let us consider case (ii). In this case, the defect supercell volume $V_D$ and the tetragonality parameter $\eta_D$ are allowed to relax in order to maintain the pressure $P_0$ and the stress $\sigma_0$. Hence, the finite-size error of the formation enthalpy, calculated in a defect supercell containing N sites can be expanded in a Taylor serier, whose first term is
\begin{eqnarray}
  \label{eq:correct}
  \tilde{H}(P_0,\sigma_0,1/N) - \tilde{H}(P_0,\sigma_0,0)  - \Delta H_F =  \nonumber \\
   \frac{c_0}{2}\frac{\partial^2 \tilde{E}}{\partial c^2} - \frac{K_L}{V_L}~\Delta V_R^2 - C_L^{\eta}~\Delta \eta_R^2 + ... ,  
\end{eqnarray}
where $K_L$ and $C_L^{\eta}$ are the bulk modulus and elastic constant of the perfect crystal. The above result is easily obtained by Taylor expansion of the generalized enthalpy $\tilde{H}(P_0,\sigma_0,1/N)$ about zero defect concentration, using the stress-strain relations Eqs.~\ref{eq:sstrain} and \ref{eq:spres}, as well as the definitions of the elastic constants Eqs.~\ref{eq:elcnst} and \ref{eq:bulk}. In this way, it can also be shown that $\Delta H_F = \Delta E_F$.

Equation~\ref{eq:correct} confirms that finite supercell-size errors introduced in calculated defect properties can often be reduced significantly whenever supecell shape and volume relaxations are allowed. The above derivation relies on the defect concentration be small enough, or supercell sizes large enough, such that corrections due to periodic image interactions can be expanded in low-order perturbation theory. This condition is usually satisfied in defect supercells exceeding 100 atoms. However, it should be noted that even in large supercells, shape and volume relaxations cannot fully account for all the periodic image interactions. They can be considered the lowest order terms in a multipole expansion of these interactions.

\bibliographystyle{unsrt}
\bibliography{ourref}


\end{document}